\documentclass[showpacs,preprintnumbers,prd,nofootinbib,floats,amssymb,floatfix]{revtex4}
\usepackage{amsmath}
\usepackage{amsfonts}
\usepackage{graphicx}
\usepackage{hyperref}
\usepackage{amsmath}
\usepackage{amstext}
 
\begin{document}

\title {Hamiltonian dynamics of  5D Kalb-Ramond theories with a compact  dimension}
 \author{ Alberto Escalante}  \email{aescalan@ifuap.buap.mx}
\affiliation{ Instituto de F{\'i}sica Luis Rivera Terrazas, Benem\'erita Universidad Aut\'onoma de Puebla, (IFUAP). Apartado
 postal J-48 72570 Puebla. Pue., M\'exico,}
 \author{Alberto  L\'opez-Villanueva} 
\affiliation{ Facultad de Ciencias F\'{\i}sico Matem\'{a}ticas, Benem\'erita Universidad Au\-t\'o\-no\-ma de Puebla,
 Apartado postal 1152, 72001 Puebla, Pue., M\'exico.}

\begin{abstract}
%We study the St{\"{u}}ckelbergâs Hamiltonians in the presence of an extra dimension 
%compactified on a circle of radius $R$ ($M^{4}\times S^{1}$ spacetime). 
A detailed Hamiltonian analysis for a five-dimensional  Kalb-Ramond, massive Kalb-Ramond and St{\"{u}}eckelberg Kalb-Ramond    theories    with a compact dimension is performed.   We develop a complete  constraint program,    then we quantize the theory by   constructing  the Dirac brackets. From the gauge transformations of the theories,   we fix  a particular gauge and we find pseudo-Goldstone bosons in    Kalb-Ramond and St{\"{u}}eckelberg Kalb-Ramond's   effective theories.  Finally we discuss some  remarks and prospects. 
\end{abstract}

\date{\today}
\preprint{}
\maketitle
\section{INTRODUCTION}
\vspace{1em} \
 It is well-know that   antisymmetric tensor fields have  an important relevance in theoretical physics. In fact, the antisymmetric tensor fields  has been used for describing  mass zero spinless as well as vector particles \cite{1, 2, 4, 4a, 5a, 6a}; in other cases, they appear  in some formulations of supergravity theories \cite{5, 6, 7} and  as a way of gauging the apparent internal supersymmetry of the weak interactions \cite{8}. In string theory, antisymmetric fields are mediators of the interaction between  open strings with charged particles \cite{9}, and also they are a fundamental block for describing the unification of Yang-Mills and supergravity \cite{10}. Moreover, they  have also an important role   characterizing defects in solid state physics \cite{11}.\\
For the reasons explained  above, in this paper  we analyze in the context of extra dimensions theories  involving antisymmetric  tensor fields. We study three models;   5D Kalb-Ramond,  5D Proca Kalb-Ramond and  5D St{\"{u}}eckelberg Kalb-Ramond    theories \cite{12}. We carryout    the compactification process on a  $S^1/\mathbf{Z_2}$ orbifold  obtaining an effective Lagrangian composed by a   four-dimensional  theory  plus a tower of $kk$-excitations. We analize the effects of the compactification process on the theory by performing a pure Dirac's framework. We develop  a complete  constraints program, we find that  5D Kalb-Ramond and 5D St{\"{u}}eckelberg Kalb-Ramond theories present reducibility conditions among the constraints in both the zero modes and in the $kk$-excitations,  while  5D Proca Kalb-Ramond is an irreducible system. We show that  5D Kalb-Ramond and 5D St{\"{u}}eckelberg Kalb-Ramond Lagrangians  are gauge theories, from the gauge transformations we fixed  the gauge and by using that gauge we obtain that there are present  pseudo-Goldstone bosons in the theories. Respect to 5D Proca Kalb-Ramond Lagrangian, the theory is not a gauge theory and there are  not present  pseudo-Goldstone bosons. Because of  5D Kalb-Ramond and 5D St{\"{u}}eckelberg Kalb-Ramond theories  are reducible systems,  we use  the phase space extension procedure  for constructing the Dirac  brackets and we calculate these  brackets among the physical fields. All these ideas  are  clarified along the paper. 
\newline
\newline
%%%%%%%%%%%%%%%%%%%%%%%%%
\section{Kalb-Ramond action in 5D with a compact dimension }
The notation that we will use along the paper is the following: the capital latin indices $M, N$ run over $0,1,2,3,5$ here $5$ label the extra compact dimension and these indices can be raised and lowered by the four-dimensional Minkowski metric $\eta_{M N}= (1,-1,-1,-1,-1)$; $y$ will represent the coordinate in the compact dimension and $\mu, \nu=0,1,2,3$ are spacetime indices, $x^\mu$  the coordinates that label the points for the four-dimensional manifold $M_4$; furthermore we will suppose that the compact dimension is a $S^1/\mathbf{Z_{2}}$ orbifold whose radius is $R$. Hence,  let us study the five dimensional Kalb-Ramond action given by \cite{12}
\begin{eqnarray}
\mathcal{L} _{} = \frac{1}{2\times 3 ! } H_{M N L} H^{ M N L},
\label{eq1}
\end{eqnarray}
where the strength fields   $H_{M N L } = \partial _{M}  B_{N L } + \partial _{N }  B_{L  M  } + \partial _{L }  B_{M N }$, with $B_{LM}=-B_{ML}$ is the Kalb-Ramond field.  In this manner, for  studying the theory in the context of Kaluza-Klein modes, we  express any dynamical variable defined on  $M_{4}\times S^1/\mathbf{Z_2}$ as a  complete set of harmonics  \cite{13, 14, 15, 16}
\begin{eqnarray}
&&B _{\mu \nu}(x, y)= \frac{1}{\sqrt{2\pi R}} B^{(0)} _{\mu \nu}(x)+  \frac{1}{\sqrt{\pi R}} \sum ^{\infty} _{n=1}  B^{(n)} _{\mu \nu}(x) \cos \Big( \frac{n y}{R} \Big), \nonumber \\
&&B _{\mu 5}(x, y)= \frac{1}{\sqrt{\pi R}} \sum ^{\infty} _{n=1}  B^{(n)} _{\mu 5}(x) \sin  \Big( \frac{n y}{R} \Big).
\label{eq4}
\end{eqnarray}
For this theory, the dynamical variables for the zero mode are given by  $B^{(0)}_{0i}, B^{(0)}_{ij}$ and for the $kk$-modes are $B^{(n)}_{0i}, B^{(n)}_{ij}, B^{(n)}_{05}, B^{(n)}_{i5}$ with   $i,j=1,2,3$. We shall  suppose that the number of $kk$-excitations  is $k$, and we will  take the limit $k \rightarrow \infty$  at the end of the calculations, thus,  $n=1, 2, 3...k-1$.\\
By taking into account  (\ref{eq4}) in (\ref{eq1})  and performing the  integration over the fifth  dimension, we obtain the following effective Lagrangian given by
\begin{eqnarray}
\mathcal{L}_{} &=& \frac{1}{2 \times 3!} H^{(0)}_{\mu \nu \lambda} H^{ \mu \nu \lambda}_{(0)} + \sum ^{\infty}_{n=1} \bigg[ \frac{1}{2 \times 3!} H^{(n)}_{\mu \nu \lambda} H^{ \mu \nu \lambda}_{(n)} \nonumber\\
&&+\frac{1}{4}\Big( \partial _{\mu } B^{(n)} _{\nu 5} + \partial _{\nu } B^{(n)} _{5 \mu} -  \frac{n}{R} B^{(n)} _{\mu \nu } \Big)
 \Big( \partial ^{\mu} B^{\nu 5} _{(n)} + \partial ^{\nu} B^{5\mu} _{(n)}  - \frac{n}{R}  B^{\mu \nu } _{(n)}  \Big) \bigg].
 \label{eq5}
\end{eqnarray}
In this manner, we can compute the following Hessian of the theory 
\begin{eqnarray}
 \frac{\partial^2 \mathcal{L}}{\partial (\partial _0 B ^{(0)}_{\lambda\rho }) \partial (\partial _0 B ^{(0)}_{\alpha \beta})}
  &=& \frac{1}{4} (g ^{\alpha  \lambda }g^{\beta \rho  }- g ^{\alpha  \rho  }g^{ \beta  \lambda}), \nonumber \\
\frac{\partial ^{2} \mathcal{L}}{ \partial (\partial _0 B ^{(m)}_{K M })  \partial (\partial _0 B ^{(h)}_{L H })} &=& \frac{1}{4}(g^{L K} g^{H M } - g^{L M} g^{H K }  )  +\frac{1}{4} \delta ^{H} _{5} \delta ^{M} _{5}
g^{L K}, 
\end{eqnarray}
it is straightforward to observe that the Hessian has a rank=4   and $4(k-1)$ null vectors, thus, we expect   $4(k-1)$ primary constraints.  Therefore, from the definition of the momenta  $ ( \Pi ^{0 i  }_{(0)}, \Pi ^{i j  }_{(0)}, \Pi ^{0 i  }_{(n)}, \Pi ^{i j  }_{(n)}, \Pi ^{0 5 }_{(n)}, \Pi ^{i 5 }_{(n)})$ canonically conjugate to $(B^{(0)}_{0i}, B^{(0)}_{ij}, B^{(n)}_{0i}, B^{(n)}_{ij}, B^{(n)}_{05}, B^{(n)}_{i5})$ given by 
\begin{eqnarray}
&&\Pi ^{0 i  }_{(0)} = 0, \quad  \Pi ^{i j  }_{(0)}= \frac{1}{2} H^{0 i j }_{(0)},  \nonumber \\
%\end{eqnarray}
%\begin{eqnarray}
&& \Pi ^{0 i }_{(n)}= 0, \quad  \Pi ^{i j}_{(n)}= \frac{1}{2}  H_{(n)}^{ 0 i j },  \quad  \Pi ^{0 5 }_{(n)}= 0, \quad \Pi ^{i 5 }_{(n)}= \frac{1}{2}( \partial ^{0} B^{ i  5} _{(n)} + \partial ^{ i } B^{5 0} _{(n)}  - \frac{n}{R}  B^{0 i } _{(n)}  ), 
\end{eqnarray}
we obtain the following $4(k-1)$ primary constraints  
\begin{eqnarray}
&& \phi ^{0 i  }_{(0)} \equiv \Pi ^{0 i  }_{(0)} \approx 0,  \nonumber \\
&&   \phi ^{0i}_{(n)} \equiv \Pi ^{0 i  }_{(n)} \approx 0, \quad \phi ^{05}_{(n)} \equiv \Pi ^{0 5  }_{(n)} \approx 0. 
\end{eqnarray}
In this manner,  by using the definition of the momenta, we obtain the following canonical Hamiltonian 
\begin{eqnarray}
H_c &=&\int d ^{3} x \bigg[ 2 B^{(0)} _{0i}  \partial _j \Pi ^{ij}_{(0)}   + \Pi^{(0)} _{i j}\Pi ^{i j}_{(0)} 
- \frac{1}{2\times 3! } H ^{(0)}_{ijk} H^{ijk}_{(0)}  +  \sum ^{\infty} _{n=1} \bigg[ 2B^{(n)} _{0i}  \partial _j \Pi ^{ij}_{(n)}    \nonumber \\
&&  + \Pi^{(n)} _{i j}\Pi ^{i j}_{(n)} 
- \frac{1}{2\times 3! } H ^{(n)}_{ijk} H^{ijk}_{(n)} 
 + 2 \Pi  ^{(n)}_ {i 5}\Pi ^{i 5}_{(n)} + 2 B ^{(n)}_{0 5 }  \partial _i \Pi ^{5i}_{(n)}  + 2\frac{n}{R} B ^{(n)}_{0 i} \Pi ^{i5}_{(n)}  \nonumber \\
 &&-\frac{1}{4}\Big( \partial _i B^{(n)} _{j 5} + \partial _j B^{(n)} _{5i} -  \frac{n}{R} B^{(n)} _{ij } \Big)
 \Big( \partial ^i B^{j5} _{(n)} + \partial ^j B^{5i} _{(n)}  -
 \frac{n}{R}  B^{ij } _{(n)}  \Big) 
\bigg] \bigg], 
\end{eqnarray}
thus, the primary Hamiltonian takes the form 
\begin{eqnarray}
H_1 = H_c + \int d ^{3}x \bigg[   
a^{(0)}_{0i} \phi ^{0i}_{(0)} + \sum _{n=1} ^{k -1} \Big( a^{(n)}_{0i} \phi ^{0i}_{(n)} +a^{(n)}_{05} \phi ^{05}_{(n)} \Big)
\bigg], 
\end{eqnarray}
where $a^{(0)}_{0i}$,  $a^{(n)}_{0i}$ and $a^{(n)}_{05} $ are Lagrange multipliers enforcing  the constraints, and the fundamental Poisson brackets are 
\begin{eqnarray}
&& \{  B _{\alpha \beta}^{(0)}(x), \Pi ^{\mu \nu} _{(0)} (z) \} = \frac{1}{2}(\delta ^{\mu}    _\alpha \delta ^{\nu} _\beta - \delta ^{\mu}    _\beta  \delta ^{\nu} _\alpha )  \delta ^{3} (x-z), \nonumber \\
&&  \{  B _{HL} ^{(l)}(x), \Pi ^{M N} _{(n)} (z) \} = \frac{1}{2}\delta ^{l} _{n}(\delta ^{M } _{H}  \delta ^{N } _{L} - \delta ^{M} _{L}  \delta ^{N } _{H}  )  \delta ^{3} (x-z).
\end{eqnarray}
Therefore, in order to determine  if there are more constraints we calculate   consistency relations among the constraints and we obtain the following secondary constraints
\begin{eqnarray}
&&\dot \phi ^{0 i } _{(0)}(x) =  \{  \phi^{0 i}_{(0)} (x), H_1 (z) \} = \partial _j \Pi ^{i j}_{(0)} (x) \approx 0,  \nonumber \\
%\end{eqnarray}
%\begin{eqnarray}
&&\dot \phi ^{0 i } _{(n)}(x) =  \{  \phi^{0 i}_{(n)} (x), H_1 (z) \} =   \partial _j \Pi ^{i j}_{(n)}(x) + \frac{n}{R}  \Pi ^{i 5}_{(n)} (x)  \approx 0,  \nonumber \\
%\end{eqnarray}
%\begin{eqnarray}
&&\dot \phi ^{0 5 } _{(n)}(x) =  \{  \phi^{0 5}_{(n)} (x), H_1 (z) \}  
=   \partial _j \Pi ^{5 j}_{(n)} (x)\approx0. %= 0 \approx \psi ^{0}(x)
\end{eqnarray}
For this theory there are not third constraints.  Therefore, we have obtained the following $8k-2$ constraints   
\begin{eqnarray}
 \phi ^{0 i  }_{(0)} & \equiv & \Pi ^{0 i  }_{(0)} \approx 0, \nonumber\\    
\psi ^{0i}_{(0)} &\equiv &  \partial _j \Pi ^{i j}_{(0)} \approx  0, \nonumber\\ 
  \phi ^{0i}_{(n)} & \equiv & \Pi ^{0 i  }_{(n)} \approx 0, \nonumber\\ \phi ^{05}_{(n)} & \equiv & \Pi ^{0 5  }_{(n)} \approx 0, \nonumber\\ 
\psi ^{0i}_{(n)} & \equiv &  \partial _j \Pi ^{i j}_{(n)} + \frac{n}{R}  \Pi ^{i 5}_{(n)} \approx  0, \nonumber \\
 \psi ^{05}_{(n)}  & \equiv &  \partial _j \Pi ^{5 j}_{(n)} \approx  0,
\end{eqnarray}
we are able to observe that these constraints are all of first class. However, they are not all independent because there are reducibility conditions among the constraints  in both,  the zero mode and the $kk$-excitations. These conditions  are given by the following $k$ relations 
\begin{eqnarray}
&&\partial _{i} \psi ^{ 0 i} _{(0)} =0,   \nonumber \\
%\end{eqnarray}
%\begin{eqnarray}
&&\partial _{i} \psi ^{ 0 i} _{(n)}  + \frac{n}{R}\psi^{05}  _{(n)}=0,
\end{eqnarray}
thus, for  the theory under study there are $[(8k-2)-k]= 7k-2$ independent first class constraints. Therefore, the counting of degrees of freedom is performed as follows; there are $20k-8$ dynamical variables and $7k-2$ independent first class constraints, thus we obtain that the number of physical degrees of freedom is given by 
\begin{eqnarray}
&&DF =\frac{1}{2}[ 20 k-8 - 2(7k-2)]= {\bf 3 k - 2},
\end{eqnarray}
we observe if  $k=1$, then there is one degree of freedom, it is  associated with the zero mode which correspond to  $4D$ Kalb-Ramond theory without an  extra dimension. \\
Because  we have obtained a set of  first class constraints,   we can calculate the gauge transformations of the theory. For this aim, we define the following gauge  generator of the theory 
\begin{eqnarray}
&&G = \int \left[  \epsilon ^{(0)}_{0i} \phi ^{0 i  }_{(0)} 
+\epsilon ^{(0)}_{i} \psi ^{0i}_{(0)} 
+    \epsilon ^{(n)}_{0i} \phi ^{0i}_{(n)} 
+ \epsilon ^{(n)}_{i} \psi ^{0i}_{(n)} 
+  \epsilon ^{(n)}_{05} \phi ^{05}_{(n)}
+  \epsilon ^{(n)}_{5}  \psi ^{05}_{(n)}\right] d^3z.  
\end{eqnarray}
In this manner, we obtain the  gauge transformations of the theory given by 
\begin{eqnarray}
&& B _{0i } ^{(0)} \rightarrow  B _{0i } ^{(0)} -  \partial _{0 } \epsilon _{i}  ^{(0)}, \nonumber \\
&&  B _{ij } ^{(0)} \rightarrow  B _{ij } ^{(0)} + \partial _{i } \epsilon _{j}  ^{(0)} - \partial _{j } \epsilon _{i}  ^{(0)} , \nonumber \\
&& 
B _{0i } ^{(n)} \rightarrow  B _{0i } ^{(n)} -  \partial _{0 } \epsilon _{i}  ^{(n)}, \nonumber \\
&& B _{05 } ^{(n)} \rightarrow  B _{05 } ^{(n)} +  \partial _{0 } \epsilon _{5}  ^{(n)}, \nonumber \\
&&  B _{ij } ^{(n)} \rightarrow  B _{ij } ^{(n)} +  \partial _{i } \epsilon _{j}  ^{(n)} - \partial _{j } \epsilon _{i}  ^{(n)} 
, \nonumber \\
&&  B _{i5 } ^{(n)} \rightarrow  B _{i5 } ^{(n)} + \frac{n}{R}  \epsilon _{i} ^{(n)} - \partial _{i}  \epsilon _{5} ^{(n)}, 
\end{eqnarray}
however, they can be written as the following compact expressions 
\begin{eqnarray}
&& \delta B _{\mu \nu  } ^{(0)} =  \partial _{\mu } \epsilon _{\nu}  ^{(0)} - \partial _{\nu  } \epsilon _{\mu}  ^{(0)},\nonumber \\
&& \delta B _{\mu \nu  } ^{(n)} =  \partial _{\mu } \epsilon _{\nu}  ^{(n)} - \partial _{\nu  } \epsilon _{\mu}  ^{(n)} , \nonumber \\
&& \delta B _{\mu 5  } ^{(n)} =  \frac{n}{R} \epsilon _{\mu}  ^{(n)} - \partial _{\mu  } \epsilon _{5}  ^{(n)},
\label{eq21a}
\end{eqnarray}
we can observe from  (\ref{eq21a}) that by fixing  the following gauge 
\begin{eqnarray}
&& \epsilon ^{(n)}_{\mu} = \frac{R}{n}( \partial _{\mu} \epsilon ^{(n)}_{5} - B_{\mu 5} ^{(n)} ),
\label{eq20}
\end{eqnarray}
we find  that  the  fields $ B _{\mu \nu  } ^{(n)}$ transforms as 
\begin{eqnarray}
&& \delta B _{\mu \nu  } ^{(n)} =  - \partial _{\mu }  B _{\nu 5  } ^{(n)} +  \partial _{\nu }  B _{\mu 5  } ^{(n)}. 
\label{eq21}
\end{eqnarray}
Therefore, by taking into account (\ref{eq20}) and (\ref{eq21}) in   the effective Lagrangian (\ref{eq5}) we obtain 
\begin{eqnarray}
\mathcal{L}_{} &=& \frac{1}{2 \times 3!} H^{(0)}_{\mu \nu \lambda} H^{ \mu \nu \lambda}_{(0)} + \sum ^{\infty}_{n=1} \bigg[ \frac{1}{2 \times 3!} H^{(n)}_{\mu \nu \lambda} H^{ \mu \nu \lambda}_{(n)} +\frac{1}{4}\Big( \frac{n}{R}\Big) ^{2}  B^{(n)} _{\mu \nu }  B^{\mu \nu } _{(n)} \bigg],
\end{eqnarray}
where we can observe that the fields $B_{\mu 5} ^{(n)} $ has been absorbed and therefore they are identified as a pseudo-Goldstone bosons. It is important to remark, that also  there are  present  pseudo-Goldstone bosons in 5D-Maxwell  and 5D-St{\"{u}}eckelberg theories  with a compact dimension \cite{14, 17}. This fact, show a close relation among Maxwell theory and Kalb-Ramond theory. \\
Now we will procedure to calculate the Dirac brackets among the physical fields. For this aim, we observe in the  constraints that there are not mixed terms of the zero modes with  the $kk$-excitations, thus, we can calculate the Dirac brackets independently for each case. First, we will calculate the Dirac brackets for the zero-mode, then for the $kk$-excitations. We need to remember that all the constraints are of  first class, hence, we need to fix the gauge in order to obtain a set of second class constraints. Because the constraints are reducible, we introduce auxiliary variables by using   the phase space extension procedure \cite{12}, thus  we will work with the following set of constraints 
\begin{eqnarray}
&&     \chi ^{1}_{(0)}  \equiv  \Pi ^{0 i  }_{(0)}, \quad  \chi ^{2}_{(0)}  \equiv  B _{0 i  } ^{(0)},  \nonumber \\
&&   \chi ^{3}_{(0)}  \equiv  2 \partial _j \Pi ^{i j}_{(0)}
+ \partial ^{i} p _{(0)} ,\quad  \chi ^{4}_{(0)}  \equiv \partial ^{j} B _{i j} ^{(0)}
+ \partial _{i} q ^{(0)},  
\label{eq25}
\end{eqnarray}
where $q_{(0)}$ y $p _{(0)}$ are auxiliary fields satisfying the following relations 
\begin{eqnarray}
 \{   q ^{(0)} (x),   p _{(0)} (z ) \}  _{} =  \delta ^{3} (x -z ). 
\end{eqnarray}
It is important to remark, that the introduction of the these auxiliary variables converts  the constraints in a set of irreducible constraints, therefore it is possible to calculate the Dirac brackets of the theory. In this way, we obtain the following matrix whose entries are the Poisson brackets among the constraints (\ref{eq25}) given by 
\begin{eqnarray}
\left(C_{\alpha \beta}^{ (0)}\right)
= \left(\begin{array}{cccc}
0 & -\frac{1}{2}\delta ^{i}_{j} & 0 & 0\\
\frac{1}{2}\delta ^{i}_{j} & 0 & 0 & 0 \\
0 & 0 & 0 &  -\delta ^{i}_{j} \nabla ^{2}  \\
0 & 0 &  \delta ^{i}_{j} \nabla ^{2} & 0
\end{array} \right)  \delta^{3}(x-z),
\end{eqnarray}
where its inverse is given by 
\begin{eqnarray}
\left(C ^{\alpha \beta}_{(0)}\right)
= \left(\begin{array}{cccc}
0 & 2\delta _{i j} & 0 & 0 \\
-2 \delta _{ij} & 0 & 0 & 0\\
 0 & 0 & 0 & \frac{ \delta ^{j}_{i}}{\nabla ^{2}}  \\
 0 & 0 &    -\frac{ \delta ^{j}_{i}}{\nabla ^{2}} & 0
\end{array} \right)  \delta^{3}(x-z).
\end{eqnarray}
In this manner, the Dirac brackets of two functionals $A$, $B$ defined on the phase space,  are expressed by
\[
\{F(x),G(z)\}_{D}\equiv\{F(x),G(z)\}+\int d^{2}ud^{2}w\{F(x),\xi_{\alpha}(u)\}C{^{\alpha\beta}}\{\xi_{\beta}(w),G(z)\},
\]
where $\{F(x),G(z)\}$ is the Poisson bracket  between two functionals $F,G$,  and $\xi_{\alpha}= (\chi_1, \chi_2, \chi_3, \chi_4)$ represent  the second class constraints. By using this fact, we obtain the following nonzero  Dirac's brackets for the zero-mode 
\begin{eqnarray}
&& \{   B _{0i} ^{(0)} (x),   \Pi ^{0j}  _{(0)} (z ) \}  _{D} =  \delta ^{j}_{i} \delta ^{3} (x -z ),\nonumber \\
%\end{eqnarray}
%\begin{eqnarray}
&& \{   B _{ij} ^{(0)} (x),   \Pi ^{kl}  _{(0)} (z ) \}  _{D} 
 =\frac{1}{2}[ \delta ^{k}_{i}\delta ^{l}_{j} - \delta ^{l}_{i}\delta ^{k}_{j} - \frac{1 }{\nabla ^2} (\delta ^{k}_{i} \partial ^{l} \partial _{j}  - \delta ^{l}_{i} \partial ^{k} \partial _{j} - \delta ^{k}_{j} \partial ^{l} \partial _{i} +  \delta ^{l}_{j} \partial ^{k} \partial _{i})  ]  \delta ^{3} (x  -z ).
\end{eqnarray}
Furthermore,  the Dirac brackets among physical and auxiliary variables  vanish 
\begin{eqnarray}
 \{   q ^{(0)} (x),   p _{(0)} (z ) \}  _{D}&=&0, \nonumber \\
 \{  q ^{(0)} (x),    \Pi ^{ij }  _{(0)} (z ) \}  _{D} &=&0, \nonumber \\
    \{  q ^{(0)} (x),    B _{ij} ^{(0)} (z ) \}  _{D} &=& 0, \nonumber \\
 \{   B _{kl} ^{(0)} (x),    p _{(0)} (z ) \}  _{D} &=&0,\nonumber \\
 \{   \Pi _{ij} ^{(0)} (x ) , p _{(0)} (z),  \}  _{D} &=& 0.
\end{eqnarray}
We are able to observe that the Dirac brackets are independent of the auxiliary variables \cite{12}. \\
Now, we will compute the Dirac brackets for the $kk$-excitations. Just as it was performed above, we fix the gauge and also we will introduce  auxiliary variables; we need to remember that  for the constraints of the $kk$-excitations there are reducibility conditions as well. In this manner, we will work  with  the following  set of independent second class constraints 
\begin{eqnarray}
\qquad && \chi ^{1}_{(n)}  \equiv  \Pi ^{0 i  }_{(n)}, \quad   \chi ^{2}_{(n)}  \equiv  B _{0 i  }^{(n)}, \nonumber \\
 &&    \chi ^{3}_{(n)}  \equiv  \Pi ^{05} _{(n)} , \quad  \chi ^{4}_{(n)}  \equiv B_{05} ^{(n)}, \nonumber \\
 &&     \chi ^{5}_{(n)}  \equiv  2 \partial _j \Pi ^{i j}_{(n)} + \frac{n}{R}2\Pi ^{i5} _{(n)} + \partial ^{i} p _{(n)} , \quad 
  \chi ^{6}_{(n)}  \equiv \partial ^j B _{i j}^{(n)} +  \partial _{i} q ^{(n)} , \nonumber \\
 &&   \chi ^{7}_{(n)}  \equiv 2 \partial _j \Pi ^{5 j}_{(n)} , \quad  \chi ^{8}_{(n)}  \equiv \partial ^j B _{5 j} ^{(n)} ,
\end{eqnarray}
just as above, the auxiliary fields   $q_{(n)}$ and  $p _{(n)}$ satisfy 
\begin{eqnarray}
 \{   q ^{(n)} (x),   p _{(n)} (z ) \}  _{} =  \delta ^{3} (x -z ). 
\end{eqnarray}
Therefore, the non-zero Poisson brackets among the constraints are given by 
\begin{eqnarray}
  \{ \chi ^{1}_{(n)}(x), \chi ^{2}_{(n)}(z) \}   & =  &  %\{ \Pi ^{0 i  }_{(n)}(x),     B _{0 j  } ^{(n)} (z)\}= 
  - \frac{1}{2}\delta _{j}^{i} \delta ^{3}(x-z),\nonumber \\
 \{ \chi ^{3}_{(n)}(x), \chi ^{4}_{(n)}(z) \}   & = & %\{ \Pi ^{0 5 }_{(n)}(x),     B _{0 5  } ^{(n)} (z)\}= 
  - \frac{1}{2} \delta ^{3}(x-z),\nonumber \\
  \{ \chi ^{5}_{(n)}(x), \chi ^{6}_{(n)}(z) \}  &=& % \{ 2 \partial _j \Pi ^{i j}_{(n)}  (x),   \partial ^l B _{k l} ^{(n)} (z) \} + \{  \partial ^{i} p _{(n)} (x),    \partial _{k} q ^{(n)}(z) \}  =
   -  \delta ^{i}_{k} \partial _{j} \partial ^{j} \delta ^{3}(x -z ), \nonumber \\
 \{ \chi ^{5}_{(n)}(x), \chi ^{8}_{(n)}(z) \}  &=&  %\{ \frac{n}{R}2\Pi ^{i5} _{(n)} (x),   \partial ^l B _{5 l} ^{(n)} (z) \} = 
 \frac{n}{R}\partial ^i \delta ^{3}(x -z ),
  \nonumber \\
 \{ \chi ^{7}_{(n)}(x), \chi ^{8}_{(n)}(z) \}  
&=& % \{ 2 \partial _j \Pi ^{5 j}_{(n)}  (x),   \partial ^{l} B _{5 l} ^{(n)} (z) \} = 
- \partial _{i} \partial ^{i} \delta ^{3}(x - z),
\end{eqnarray}
thus, we obtain the following matrix 
\begin{eqnarray}
\left(C_{\alpha \beta} ^{(n)}\right)
= \left(\begin{array}{cccccccc}
 0 & -\frac{1}{2}\delta ^{i}_{j} & 0 & 0 & 0 & 0 & 0 & 0\\
 \frac{1}{2}\delta ^{i}_{j} & 0 & 0 & 0 & 0 & 0 & 0 & 0 \\
 0 & 0 & 0 & -\frac{1}{2} & 0 & 0 & 0 & 0 \\
 0 & 0 &  \frac{1}{2} & 0 & 0 & 0 & 0 & 0 \\
 0 & 0 & 0 & 0 & 0 & -\delta ^{i}_{j}\nabla ^{2} & 0 & \frac{n}{R}\partial ^i  \\
 0 & 0 & 0 & 0 & \delta ^{i}_{j}\nabla ^{2} & 0 & 0 & 0 \\
 0 & 0 & 0 & 0 & 0 & 0 & 0 &  -\nabla ^{2}\\
0 & 0  & 0 & 0 & -\frac{n}{R}\partial ^i & 0 & \nabla ^{2} & 0 
\end{array} \right)  \delta^{3}(x-z), \nonumber 
\end{eqnarray}
where its inverse is given by 
\begin{eqnarray}
\left(C^{\alpha \beta} _{(n)}\right)
= \left(\begin{array}{cccccccc}
 0 & 2\delta ^{j}_{i} & 0 & 0 & 0 & 0  & 0 & 0 \\
 -2 \delta ^{j}_{i} & 0 & 0 & 0 & 0 & 0 & 0 & 0 \\
 0 & 0 & 0 & 2 & 0 & 0 & 0 & 0  \\
 0 & 0 & -2 & 0 & 0 & 0 & 0 & 0  \\
0 & 0 & 0 & 0 & 0 & \frac{ \delta ^{j}_{i}} {\nabla ^{2}} & 0 & 0 \\
 0 & 0 & 0 & 0 & -\frac{ \delta ^{j}_{i}}{\nabla ^{2}} & 0 & -\frac{n \partial ^{j}}{R(\nabla ^{2})^2} & 0 \\
0 & 0 & 0 & 0 & 0 & \frac{n \partial ^{j}}{R(\nabla ^{2})^2} & 0 &  \frac{1}{\nabla ^{2}} \\
 0 & 0 & 0 & 0 & 0 & 0 &  -\frac{1}{\nabla ^{2}} & 0 
\end{array} \right)  \delta^{3}(x-z). \nonumber
\end{eqnarray}
In this way, we obtain the following non-zero  Dirac brackets 
\begin{eqnarray}
&& \{   B _{0i} ^{(n)} (x),   \Pi ^{0j}  _{(n)} (z ) \}  _{D}  =  \delta ^{j}_{i} \delta ^{3} (x -z ),\nonumber \\
%\end{eqnarray}
%\begin{eqnarray}
 && \{   B _{ij} ^{(n)} (x),   \Pi ^{kl}  _{(n)} (z ) \}  _{D} 
 =\frac{1}{2}[ \delta ^{k}_{i}\delta ^{l}_{j} - \delta ^{l}_{i}\delta ^{k}_{j} - \frac{1 }{\nabla ^2} (\delta ^{k}_{i} \partial ^{l} \partial _{j}  - \delta ^{l}_{i} \partial ^{k} \partial _{j} - \delta ^{k}_{j} \partial ^{l} \partial _{i} +  \delta ^{l}_{j} \partial ^{k} \partial _{i})  ]  \delta ^{3} (x  -z ).\quad 
\end{eqnarray}
and the Dirac brackets between  physical and auxiliary variables  vanish as expected, this is 
\begin{eqnarray}
 \{   q ^{(n)} (x),   p _{(n)} (z ) \}  _{D}  &=&0, \nonumber \\
 \{  q ^{(n)} (x),    \Pi ^{ij }  _{(n)} (z ) \}  _{D} & =&0,\nonumber \\
  \{  q ^{(n)} (x),    B _{ij} ^{(n)} (z ) \}  _{D} &=& 0,\nonumber \\
 \{   B _{kl} ^{(n)} (x),    p _{(n)} (z ) \}  _{D} &=&0, \nonumber \\
\{   \Pi _{ij} ^{(n)} (x ) , p _{(n)} (z),  \}  _{D} &=& 0.
\end{eqnarray}
%%%%%%%%%%%%%%%%%%%%%%%%%%%%%%
%%%%%%%%%%%%%%%%%%%%%%%%%%%%%%%%%%%%%%%%%%%%%
\section{ 5D Proca Kalb-Ramond theory with a compact dimension}
In this section we shall analyze the following action  
\begin{eqnarray}
\mathcal{L} _{} = \frac{1}{2\times 3 ! } H_{M N L} H^{ M N L} -  \frac{1}{4}m^{2} B_{M N }  B^{M N },
\label{eq46}
\end{eqnarray}
where the fields $B_{MN}$ and $H^{MNK}$ are defined as above. By performing the 4+1 decomposition, the Lagrangian (\ref{eq46}) takes the form
\begin{eqnarray}
\mathcal{L}_{} &=&  \frac{1}{2 \times 3!} H^{}_{\mu \nu \lambda} H^{ \mu \nu \lambda}_{} +\frac{1}{4} H^{}_{5 \mu \nu } H^{ 5 \mu \nu}_{} 
- \frac{1}{4}m^{2} B ^{}_{\mu \nu  }  B^{\mu \nu  }_{} 
- \frac{1}{2}m^{2} B ^{}_{ \mu 5  }  B^{ \mu 5  }_{},
\end{eqnarray}
thus, by taking  into account the expansion (\ref{eq4})  and integrating over the compact  dimension we obtain the following effective Lagrangian 
\begin{eqnarray}
\mathcal{L}_{} &=& \frac{1}{2 \times 3!} H^{(0)}_{\mu \nu \lambda} H^{ \mu \nu \lambda}_{(0)} - \frac{1}{4}m^{2} B ^{(0)}_{\mu \nu  }  B^{\mu \nu  }_{(0)}  + \sum ^{\infty}_{n=1} \bigg[ \frac{1}{2 \times 3!} H^{(n)}_{\mu \nu \lambda} H^{ \mu \nu \lambda}_{(n)}  - \frac{1}{4}m^{2}  B ^{(n)}_{\mu \nu  }  B^{\mu \nu  }_{(n)}  \nonumber\\
&&-\frac{1}{2}m^{2} B^{(n)} _{\mu 5} B_{(n)} ^{\mu 5}  
+\frac{1}{4}\Big( \partial _{\mu } B^{(n)} _{\nu 5} + \partial _{\nu } B^{(n)} _{5 \mu} -  \frac{n}{R} B^{(n)} _{\mu \nu } \Big)
 \Big( \partial ^{\mu} B^{\nu 5} _{(n)} + \partial ^{\nu} B^{5\mu} _{(n)}  - \frac{n}{R}  B^{\mu \nu } _{(n)}  \Big) \bigg]. 
\end{eqnarray}
In order to perform the Hamiltonian analysis, we observe that the Hessian 
\begin{eqnarray}
&&\frac{\partial^2 \mathcal{L}}{\partial (\partial _0 B ^{(0)}_{\lambda\rho }) \partial (\partial _0 B ^{(0)}_{\alpha \beta})}
 = \frac{1}{4} (g ^{\alpha  \lambda }g^{\beta \rho  }- g ^{\alpha  \rho  }g^{ \beta  \lambda}), \nonumber \\
%\end{eqnarray}
%\begin{eqnarray}
&&\frac{\partial ^{2} \mathcal{L}}{ \partial (\partial _0 B ^{(m)}_{K M })  \partial (\partial _0 B ^{(h)}_{L H })}=\frac{1}{4}(g^{L K} g^{H M } - g^{L M} g^{H K }  )  +\frac{1}{4} \delta ^{H} _{5} \delta ^{M} _{5}
g^{L K},
\end{eqnarray}
has a rank=4   and $4(k-1)$ null vectors, thus, we expect   $4(k-1)$ primary constraints.  Therefore, from the definition of the momenta  $ ( \Pi ^{0 i  }_{(0)}, \Pi ^{i j  }_{(0)}, \Pi ^{0 i  }_{(n)}, \Pi ^{i j  }_{(n)}, \Pi ^{0 5 }_{(n)}, \Pi ^{i 5 }_{(n)})$ canonically conjugate to $(B^{(0)}_{0i}, B^{(0)}_{ij}, B^{(n)}_{0i}, B^{(n)}_{ij}, B^{(n)}_{05}, B^{(n)}_{i5})$ we obtain 
\begin{eqnarray}
&&\Pi ^{0 i  }_{(0)} = 0, \quad  \Pi ^{i j  }_{(0)}= \frac{1}{2} H^{0 i j }_{(0)},  \nonumber \\
&& \Pi ^{0 i }_{(n)}= 0, \quad   \Pi ^{i j}_{(n)}= \frac{1}{2}  H_{(n)}^{ 0 i j } , \quad    \Pi ^{0 5 }_{(n)}= 0, \quad 
\Pi ^{i 5 }_{(n)}= \frac{1}{2}( \partial ^{0} B^{ i  5} _{(n)} + \partial ^{ i } B^{5 0} _{(n)}  - \frac{n}{R}  B^{0 i } _{(n)}  ), 
\end{eqnarray}
thus, we identify  the following $4k-1$ primary constraints 
\begin{eqnarray}
&& \phi ^{0 i  }_{(0)} \equiv \Pi ^{0 i  }_{(0)} \approx 0,   \nonumber \\
&&  
 \phi ^{0i}_{(n)} \equiv \Pi ^{0 i  }_{(n)} \approx 0, \quad 
 \phi ^{05}_{(n)} \equiv \Pi ^{0 5  }_{(n)} \approx 0.
\end{eqnarray}
%%%%%%%%%%%%%%%%%%%%%%%%%%%%%%%%%%%%%%%%%%%%%%%%%%%%%%
By using the definition of the momenta, we obtain the canonical Hamiltonian
\begin{eqnarray}
H_c  &=&\int d ^{3} x \bigg[ 2B^{(0)} _{0i}  \partial _j \Pi ^{ij}_{(0)} + \Pi^{(0)} _{i j}\Pi ^{i j}_{(0)} 
- \frac{1}{2\times 3! } H ^{(0)}_{ijk} H^{ijk}_{(0)} + \frac{1}{2} m^{2} B ^{(0)}_{0i }  B^{0i }_{(0)} + \frac{1}{4} m^{2} B ^{(0)}_{ij }  B^{ij }_{(0)} \nonumber \\
&& %+ \frac{1}{4} m^{2} B ^{(0)}_{ij }  B^{ij }_{(0)} 
+ \sum ^{\infty} _{n=1} \bigg[ 2B^{(n)} _{0i}  \partial _j \Pi ^{ij}_{(n)}   + \Pi^{(n)} _{i j}\Pi ^{i j}_{(n)} 
- \frac{1}{2\times 3! } H ^{(n)}_{ijk} H^{ijk}_{(n)} + \frac{1}{2} m^{2} B ^{(0)}_{0i }  B^{0i }_{(0)} + \frac{1}{4} m^{2}  B ^{(n)}_{ij }  B^{ij }_{(n)} \nonumber \\
&& %+ \frac{1}{4} m^{2}  B ^{(n)}_{ij }  B^{ij }_{(n)}
+ \frac{1}{2}m^{2} B^{(n)} _{05} B_{(n)} ^{05} + \frac{1}{2}m^{2}  B ^{(n)} _{i5}  B_{(n)} ^{i5} +
  2 \Pi  ^{(n)}_ {i 5}\Pi ^{i 5}_{(n)} + 2B ^{(n)}_{0 5 }  \partial _i \Pi ^{5i}_{(n)} + \frac{n}{R}2 B ^{(n)}_{0 i} \Pi ^{i5}_{(n)} \nonumber \\
&&%+ \frac{n}{R}2 B ^{(n)}_{0 i} \Pi ^{i5}_{(n)}   
-\frac{1}{4}\Big( \partial _i B^{(n)} _{j 5} + \partial _j B^{(n)} _{5i} -  \frac{n}{R} B^{(n)} _{ij } \Big)
 \Big( \partial ^i B^{j5} _{(n)} + \partial ^j B^{5i} _{(n)}  -
 \frac{n}{R}  B^{ij } _{(n)}  \Big) 
\bigg] \bigg] ,
\end{eqnarray}
and the primary Hamiltonian is given by 
\begin{eqnarray}
H_1 = H_c + \int d ^{3}x \bigg[ a^{(0)}_{0i} \phi ^{0i}_{(0)} + \sum _{n=1} ^{k - 1} \Big( a^{(n)}_{0i} \phi ^{0i}_{(n)} +a^{(n)}_{05} \phi ^{05}_{(n)} \Big)
\bigg],
\end{eqnarray}
where $a^{(0)}_{0i}$,  $a^{(n)}_{0i}$ and $a^{(n)}_{05} $ are Lagrange multipliers enforcing  the constraints. The fundamental Poisson brackets of the theory are as usual 
\begin{eqnarray}
&&  \{  B _{\alpha \beta}^{(0)}(x), \Pi ^{\mu \nu} _{(0)} (z) \} = \frac{1}{2}(\delta ^{\mu}    _\alpha \delta ^{\nu} _\beta - \delta ^{\mu}    _\beta  \delta ^{\nu} _\alpha )  \delta ^{3} (x-z), \nonumber \\
&&  \{  B _{HL} ^{(l)}(x), \Pi ^{M N} _{(n)} (z) \} = \frac{1}{2}\delta ^{l} _{n}(\delta ^{M } _{H}  \delta ^{N } _{L} - \delta ^{M} _{L}  \delta ^{N } _{H}  )  \delta ^{3} (x-z).
\end{eqnarray}
In oder to observe if there are more constraints, we demand consistency conditions  for the primary constraints and we obtain the following secondary constraints
\begin{eqnarray}
&& \dot \phi ^{0 i } _{(0)}(x) =  \{  \phi^{0 i}_{(0)} (x), H_1 (z) \}  
=  2 \partial _j \Pi ^{i j}_{(0)} (x) + m^{2} B^{ 0 k } _{(0)}(x) \approx 0, \nonumber \\
%\end{eqnarray}
%%%%%%%%%%%%%%%%%%%%%%%%%%%%%%%%%%
%\begin{eqnarray}
&& \dot \phi ^{0 i } _{(n)}(x) =  \{  \phi^{0 i}_{(n)} (x), H_1 (z) \} 
 =  2 \partial _j \Pi ^{i j}_{(n)} (x) + m ^2 B^{ 0 i }_{(n)} (x) + \frac{n}{R} 2 \Pi ^{i 5}_{(n)} (x)  \approx 0, \nonumber \\
%\end{eqnarray}
%\begin{eqnarray}
&& \dot \phi ^{0 5 } _{(n)}(x) =  \{  \phi^{0 5}_{(n)} (x), H_1 (z) \}   
= 2 \partial _j \Pi ^{5 j}_{(n)} (x) + m^{2} B_{(n)} ^{05} (x)\approx 0, 
\end{eqnarray}
for this theory there are not third constraints. Therefore, the full set of constraints for the theory is given by 
\begin{eqnarray}
\qquad && \phi ^{0 i  }_{(0)} \equiv \Pi ^{0 i  }_{(0)} \approx 0,\nonumber \\  
&& \psi ^{0i}_{(0)} \equiv  2 \partial _j \Pi ^{i j}_{(0)} + m^{2}B^{ 0 i } _{(0)}  \approx  0,\nonumber \\
&& \phi ^{0i}_{(n)} \equiv \Pi ^{0 i  }_{(n)} \approx 0, \nonumber \\
&& \psi ^{0i}_{(n)}  \equiv  2 \partial _j \Pi ^{i j}_{(n)} + m^{2}B^{ 0 i } _{(n)} + \frac{n}{R} 2 \Pi ^{i 5}_{(n)} \approx  0,\nonumber \\
&& \phi ^{05}_{(n)} \equiv \Pi ^{0 5  }_{(n)} \approx 0,\nonumber \\
&& \psi ^{05}_{(n)}   \equiv  2 \partial _j \Pi ^{5 j}_{(n)} + m^{2} B_{(n)} ^{05} \approx  0.
\end{eqnarray}
We  can observe that the constraints given above are of second class and there are not reducibility conditions. In fact, the term of mass  breaks down both,  the gauge invariance of the kinetic term and  the reducibility conditions among the constraints. Therefore, the counting of physical degrees of freedom is carry out in the following form; there are  $20k-8$ dynamical variables and $8k-2$ independent second class constraints, thus there are  
\begin{eqnarray}
DF &=&\frac{1}{2}[ 20k-8 -(8k-2)]
= {\bf 6 k - 3}
\end{eqnarray}
degrees of freedom. We observe that if we take $k=1$, then we obtain  $DF ={\bf 3}$ as expected. On the other hand, we can observe that  each excitation  contribute with 6 degrees of freedom. \\
Now we will calculate  the Dirac brackets of the theory. For this aim, we rewrite the constraints in the following form 
\begin{eqnarray}
&& \chi ^{1  }_{(0)} \equiv \Pi ^{0 i  }_{(0)}, \quad   \chi  ^{2}_{(0)} \equiv  2 \partial _j \Pi ^{i j}_{(0)} + m^{2}B^{ 0 i } _{(0)} , \nonumber \\ &&   \chi ^{1}_{(n)} \equiv \Pi ^{0 i  }_{(n)} , \quad  \chi ^{2}_{(n)}  \equiv  2 \partial _j \Pi ^{i j}_{(n)} + m^{2}B^{ 0 i } _{(n)} + \frac{n}{R} 2 \Pi ^{i 5}_{(n)},  \quad 
  \chi ^{3}_{(n)} \equiv \Pi ^{0 5  }_{(n)} , \quad   \chi ^{4}_{(n)}   \equiv  2 \partial _j \Pi ^{5 j}_{(n)} + m^{2} B_{(n)} ^{05} , \nonumber \quad 
\end{eqnarray}
we observe that the zero-modes and the excited modes are not mixed in the constraints, hence, we will calculate the Dirac brackets independently as was performed in above section. For the zero-mode we obtain 
\begin{eqnarray}
&& \{  \chi ^{1}_{(0)} (x), \chi ^{2}_{(0)} (z)\}  =  %\{  \Pi ^{0 l  }_{(0)} (x), - m^{2}B_{ 0 i } ^{(0)} (z) \}  =  
\frac{1}{2} m^{2} \delta ^{l}_{i} \delta ^{3}(x - z ), 
\end{eqnarray}
thus, the matrix whose entries are the Poisson brackets among the second class constraints for the zero-mode  take the form 
\begin{eqnarray*}
\left(C_{\alpha \beta}^{ (0)}\right)%\left( \begin{array}{cc}
% \{ \phi_{}^{1}(x),\phi_{}^{1}(z)\}  &  \{ \phi_{}^{1}(x),\phi_{}^{2}(z)\}  \\ 
% \{ \phi_{}^{2}(x),\phi_{}^{1}(z)\}  &  \{ \phi_{}^{2}(x),\phi_{}^{2}(z)\} 
%\end{array} \right) \\
= \left(\begin{array}{cc}
0 & 1 \\ 
- 1 & 0
\end{array} \right) \frac{1}{2} m^{2}\delta ^{i}_{j}  \delta^{3}(x-y),
\end{eqnarray*}
and it has an inverse given by  
\begin{eqnarray*}
\left(C^{\alpha  \beta}_{ (0)}\right)
=\left(\begin{array}{cc}
0 & - 1 \\ 
 1 & 0
\end{array} \right) \frac{2}{m^{2}}  \delta _{ij }  \delta^{3}(x-y).
\end{eqnarray*}
In this manner, the Dirac brackets of two functionals $A$, $B$ defined on the phase space,  is expressed by
\[
\{F(x),G(z)\}_{D}\equiv\{F(x),G(z)\}+\int d^{2}ud^{2}w\{F(x),\xi_{\alpha}(u)\}C{^{\alpha\beta}}\{\xi_{\beta}(w),G(z)\},
\]
where $\{F(x),G(z)\}$ is the Poisson bracket  between two functionals $F,G$,  and $\xi_{\alpha}= (\chi_1, \chi_2)$ represent  the second class constraints. By using this fact, we obtain the following  nonzero Dirac's brackets for the zero-mode \begin{eqnarray}
 \{   B_{ 0 i  } ^{(0)} (x),  B_{ pq  } ^{(0)} (z ) \}  _{D} 
 &=&- \frac{1}{m^{2}} (\delta _{ i p} \delta ^{j}_{q} - \delta _{i q} \delta ^{j}_{q} ) \partial _{j} \delta ^{3} (x  - z )\nonumber \\
 \{   B_{ 0 i  } ^{(0)} (x),  \Pi ^{0 q  } _{(0)} (z )   \}  _{D} &=&  \delta ^{q}_{i}  \delta ^{3} (x  -z ).
\end{eqnarray}
Now, we will calculate the Dirac brackets for the $kk$-excitations. For this aim, we calculate the Poisson brackets among the second class constraints of the $kk$-excitations. The nonzero brackets are given by 
\begin{eqnarray}
&& \{  \chi ^{1}_{(n)} , \chi ^{2}_{(n)} \}  = % \{  \Pi ^{0 l  }_{(n)}, - m^{2}B_{ 0 i } ^{(n)} \}  =  
\frac{1}{2} m^{2} \delta ^{l}_{i} \delta ^{3}(x - z ), \nonumber \\
 &&  \{  \chi ^{3}_{(n)} , \chi ^{4}_{(n)} \}  = % \{   \Pi ^{0 5  }_{(n)} , - m^{2}B_{ 0 5 } ^{(n)} \}   =  
 \frac{1}{2} m^{2} \delta ^{3}(x - z ), 
\end{eqnarray}
thus, the matrix whose entries are the poisson brackets among the second class constraints is given by 
\begin{eqnarray*}
\left(C_{\alpha \beta}^{ (n)}\right)%\left( \begin{array}{cc}
% \{ \phi_{}^{1}(x),\phi_{}^{1}(z)\}  &  \{ \phi_{}^{1}(x),\phi_{}^{2}(z)\}  \\ 
% \{ \phi_{}^{2}(x),\phi_{}^{1}(z)\}  &  \{ \phi_{}^{2}(x),\phi_{}^{2}(z)\} 
%\end{array} \right) \\
= \left(\begin{array}{cccc}
0 & \delta ^{i}_{l} & 0 & 0 \\ 
-\delta ^{i}_{l} & 0 & 0 & 0 \\
0 & 0 & 0 & 1 \\
0 & 0 & -1 & 0 
\end{array} \right) \frac{1}{2} m^{2} \delta^{3}(x-z),
\end{eqnarray*}
this matrix has as inverse 
\begin{eqnarray*}
\left(C^{\alpha \beta}_{ (n)}\right)
=\left(\begin{array}{cccc}
0 & - \delta _{i}^{l} & 0 & 0 \\ 
 \delta _{i}^{l} & 0 & 0 & 0 \\
0 & 0 & 0 & -1 \\
0 & 0 & 1 & 0 
\end{array} \right) \frac{2}{m^{2}}   \delta^{3}(x-z).
\end{eqnarray*}
In this manner,  we obtain the following nonzero Dirac brackets for the $kk$-excitations 
\begin{eqnarray}
&& \{   B_{ 0 i  } ^{(n)} (x),  B_{ pq  } ^{(n)} (z ) \}  _{D} = - \frac{1}{m^{2}} (\delta _{ i p} \delta ^{j}_{q} - \delta _{i q} \delta ^{j}_{q} ) \partial _{j} \delta ^{3} (x  - z ), \nonumber \\
%\end{eqnarray}
%\begin{eqnarray}
&& \{   B_{ 0 i  } ^{(n)} (x),  \Pi ^{0 q  } _{(n)} (z )   \}  _{D} 
 =  \delta ^{q}_{i}  \delta ^{3} (x  -z ),\nonumber \\
%\end{eqnarray}
%\begin{eqnarray}
&& \{   B_{ 0 i  } ^{(n)} (x),  B_{ q 5 }^{(n)} (z )   \}  _{D}  = \frac{n}{R m^{2}} \delta _{i q}  \delta ^{3} (x  -z ).
\end{eqnarray}
Therefore, we have computed  the Dirac brackets of the theory and we can perform its canonical quantization.  
%%%%%%%%%%%%%%%%%%%%%%%%
%%%%%%%%%%%%%%%%%%%%%%%%%%
\section{5D St{\"{u}}eckelberg  Kalb-Ramond theory  with a compact dimension}
Now,  we will study the following action 
\begin{eqnarray}
\mathcal{L} _{} = \frac{1}{2\times 3 ! } H_{M N L} H^{ M N L} - \frac{1}{4}(m B_{M N } - \Phi _{M N})(m B^{M N } - \Phi ^{M N}),
\label{eq68}
\end{eqnarray}
where the field strength  $H_{M N L } $ are defined as above, $ \Phi_N$ is the St{\"{u}}eckelberg  field  and  $\Phi _{M N }= \partial _M \Phi_N  - \partial _N  \Phi_M $ \cite{12}. 
Just as in  above sections,  we can expand the fields in terms of the following series 
\begin{eqnarray}
&&\Phi _{\mu}(x, y)= \frac{1}{\sqrt{2\pi R}} \Phi ^{(0)}_{\mu}(x)+  \frac{1}{\sqrt{\pi R}} \sum ^{\infty} _{n=1}  \Phi ^{(n)}_{\mu}(x)\cos \Big( \frac{n y}{R} \Big), \nonumber \\
&&\Phi _{5}(x, y)=  \frac{1}{\sqrt{\pi R}} \sum ^{\infty} _{n=1}  \Phi ^{(n)}_5(x)\sin  \Big( \frac{n y}{R} \Big), \nonumber \\
&&B _{\mu \nu}(x, y)= \frac{1}{\sqrt{2\pi R}} B^{(0)} _{\mu \nu}(x)+  \frac{1}{\sqrt{\pi R}} \sum ^{\infty} _{n=1}  B^{(n)} _{\mu \nu}(x) \cos \Big( \frac{n y}{R} \Big), \nonumber \\
&&B _{\mu 5}(x, y)= \frac{1}{\sqrt{\pi R}} \sum ^{\infty} _{n=1}  B^{(n)} _{\mu 5}(x) \sin  \Big( \frac{n y}{R} \Big).
\label{eq69}
\end{eqnarray}
By performing the 4+1 decomposition in the Lagrangian  (\ref{eq68}),  taking into account  the series (\ref{eq69}) and integrating over  the fifth  dimension,  we obtain the following effective Lagrangian 
\begin{eqnarray}
\mathcal{L}_{} &=& \frac{1}{2 \times 3!} H^{(0)}_{\mu \nu \lambda} H^{ \mu \nu \lambda}_{(0)} - \frac{1}{4}(m B ^{(0)}_{\mu \nu  } - \Phi^{(0)} _{\mu \nu })(m B^{\mu \nu  }_{(0)} - \Phi ^{\mu  \nu }_{(0)})  \nonumber\\
&& + \sum ^{\infty}_{n=1} \bigg[ \frac{1}{2 \times 3!} H^{(n)}_{\mu \nu \lambda} H^{ \mu \nu \lambda}_{(n)}  - \frac{1}{4}(m B ^{(n)}_{\mu \nu  } - \Phi^{(n)} _{\mu \nu })(m B^{\mu \nu  }_{(n)} - \Phi ^{\mu  \nu }_{(n)}) \nonumber\\
&& -\frac{1}{2}\Big( mB^{(n)} _{\mu 5} - \partial _{\mu } \Phi^{(n)} _5 -   \frac{n}{R}  \Phi^{(n)} _{\mu} \Big) \Big( mB_{(n)} ^{\mu 5} - \partial ^{\mu } \Phi ^{5} _{(n)}  -  \frac{n}{R}\Phi _{(n)} ^{\mu} \Big) \nonumber\\
&&+\frac{1}{4}\Big( \partial _{\mu } B^{(n)} _{\nu 5} + \partial _{\nu } B^{(n)} _{5 \mu} -  \frac{n}{R} B^{(n)} _{\mu \nu } \Big)
 \Big( \partial ^{\mu} B^{\nu 5} _{(n)} + \partial ^{\nu} B^{5\mu} _{(n)}  - \frac{n}{R}  B^{\mu \nu } _{(n)}  \Big) \bigg].
 \label{eq70}
\end{eqnarray}
We can observe that the effective Lagrangian describes a $4D$ St{\"{u}}eckelberg  Kalb-Ramond theory  plus a tower of $kk$-excitations.  For this theory, the Hessian of the Lagrangian  (\ref{eq70}) given by 
\begin{eqnarray}
&& \frac{\partial ^2 \mathcal{L}}{\partial (\partial _0 \Phi ^{(l)}_ M  )\partial (\partial _0 \Phi ^{(l)}_ L  ) } 
 =( g^{L 0 } g ^{ 0 M } -   g^{L M  })  + \delta ^{M} _{5} \delta ^{L} _{5} ,\nonumber \\
%\end{eqnarray} 
%\begin{eqnarray}
&&\frac{\partial ^{2} \mathcal{L}}{ \partial (\partial _0 B ^{(m)}_{K M })  \partial (\partial _0 B ^{(h)}_{L H })} = \frac{1}{4}(g^{L K} g^{H M } - g^{L M} g^{H K }  )  +\frac{1}{4} \delta ^{H} _{5} \delta ^{M} _{5}
g^{L K},
\end{eqnarray}
has rank$= 8k-7$ and $5k-1$ null vectors, this means that we expect $5k-1$ primary constraints. Therefore, from the definition of the momenta  $ (\Pi^0_{(0)}, \Pi^i_{(0)},  \Pi ^{0 i  }_{(0)}, \Pi ^{i j  }_{(0)}, \Pi ^{0 i  }_{(n)}, \Pi ^{i j  }_{(n)}, \Pi ^{0 5 }_{(n)}, \Pi ^{i 5 }_{(n)}, \Pi^5_{(n)}, \Pi^i_{(n)}, \Pi ^{0}_{(n)})$ canonically conjugate to $(\Phi_0^{(0)}, \Phi_i ^{(0)}, B^{(0)}_{0i}, B^{(0)}_{ij}, B^{(n)}_{0i}, B^{(n)}_{ij}, B^{(n)}_{05}, B^{(n)}_{i5}, \Phi_5^{(n)}, \Phi_{i}^{(n)}, \Phi _0 ^{(n)})$ we obtain 
\begin{eqnarray}
&& \Pi ^{0 }_{(0)} = 0, \quad \Pi ^{i }_{(0)}= m B^{0 i }_{(0)} - \Phi ^{0 i } _{(0)},  \quad 
\Pi ^{0 i  }_{(0)} = 0, \quad \Pi ^{i j  }_{(0)}= \frac{1}{2} H^{0 i j }_{(0)}, \nonumber \\
&& \Pi ^{0}_{(n)}=0, \quad \Pi ^{i }_{(n)}=  m B _{(n)} ^{0 i }- \Phi ^{0 i  } _{(n)},  \quad \Pi ^{5 }_{(n)}=  mB_{(n)} ^{05} - \partial ^0 \Phi ^{5} _{(n)} - \frac{n}{R}\Phi^{0}_{(n)}, \nonumber \\
&& \Pi ^{0 i }_{(n)}= 0, \quad \Pi ^{0 5 }_{(n)}= 0, \quad \Pi ^{i j}_{(n)}= \frac{1}{2}  H_{(n)}^{ 0 i j }, \quad \Pi ^{i 5 }_{(n)}= \frac{1}{2}( \partial ^{0} B^{ i  5} _{(n)} + \partial ^{ i } B^{5 0} _{(n)}  - \frac{n}{R}  B^{0 i } _{(n)}  ),
\end{eqnarray}
thus, we identify the following $5k-1$ primary constraints 
\begin{eqnarray}
&& \phi ^{0}_{(0)}\equiv \Pi ^{0 }_{(0)} \approx 0, \quad  \phi ^{0 i  }_{(0)} \equiv \Pi ^{0 i  }_{(0)} \approx 0, \nonumber \\
&& \phi ^{0}_{(n)} \equiv \Pi ^{0}_{(n)} \approx 0,  \quad \phi ^{0i}_{(n)} \equiv \Pi ^{0 i  }_{(n)} \approx 0, \quad \phi ^{05}_{(n)} \equiv \Pi ^{0 5  }_{(n)} \approx 0. \quad
\end{eqnarray}
On the other hand,  by using the definition of the momenta we identify the  canonical Hamiltonian given by 
\begin{eqnarray}
H_c &=&\int d ^{3} x \bigg[ B^{(0)} _{0i} (m \Pi ^i_{(0)} + 2 \partial _j \Pi ^{ij}_{(0)} ) -  \Phi^{(0)} _0 \partial _i \Pi ^i _{(0)} -\frac{1}{2} \Pi^{(0)} _ i \Pi ^i_{(0)} + \Pi^{(0)} _{i j}\Pi ^{i j}_{(0)} 
- \frac{1}{2\times 3! } H ^{(0)}_{ijk} H^{ijk}_{(0)}  \nonumber \\
&& + \frac{1}{4}(m B ^{(0)}_{ij } - \Phi^{(0)} _{ij})(m B^{ij }_{(0)} - \Phi ^{ij}_{(0)})+ \sum ^{\infty} _{n=1} \bigg[B^{(n)} _{0i} (m \Pi ^i_{(n)} + 2 \partial _j \Pi ^{ij}_{(n)} ) -  \Phi^{(n)} _0 \partial _i \Pi ^i _{(n)}  \nonumber \\
&& -\frac{1}{2} \Pi^{(n)} _ i \Pi ^i_{(n)} + \Pi^{(n)} _{i j}\Pi ^{i j}_{(n)} 
- \frac{1}{2\times 3! } H ^{(n)}_{ijk} H^{ijk}_{(n)} + \frac{1}{4}(m B ^{(n)}_{ij } - \Phi^{(n)} _{ij})(m B^{ij }_{(n)} - \Phi ^{ij}_{(n)})\nonumber \\
&&
- \frac{1}{2}\Pi ^{(n)}_5  \Pi ^{5}_{(n)} +  2 \Pi  ^{(n)}_ {i 5}\Pi ^{i 5}_{(n)} + B ^{(n)}_{0 5 } (m \Pi ^{5}_{(n)} +  2 \partial _i \Pi ^{5i}_{(n)}) + \frac{n}{R}(2 B ^{(n)}_{0 i} \Pi ^{i5}_{(n)}      -\Phi^{(n)} _ 0 \Pi ^{5}_{(n)} )\nonumber \\
 &&+\frac{1}{2}\Big( mB^{(n)} _{i5} - \partial _i \Phi^{(n)} _5 -  \frac{n}{R} \Phi^{(n)} _i \Big)
 \Big( mB_{(n)} ^{i5} - \partial ^i \Phi ^{5} _{(n)}  - \frac{n}{R}  \Phi _{(n)} ^i \Big)\nonumber \\
 &&-\frac{1}{4}\Big( \partial _i B^{(n)} _{j 5} + \partial _j B^{(n)} _{5i} -  \frac{n}{R} B^{(n)} _{ij } \Big)
 \Big( \partial ^i B^{j5} _{(n)} + \partial ^j B^{5i} _{(n)}  -
 \frac{n}{R}  B^{ij } _{(n)}  \Big) 
\bigg] \bigg] ,
\end{eqnarray}
and the primary Hamiltonian takes the following form 
\begin{eqnarray}
H_1 = H_c + \int d ^{3}x \bigg[   
a^{(0)}_{0} \phi ^{0}_{(0)}  + a^{(0)}_{0i} \phi ^{0i}_{(0)} + \sum _{n=1} ^{k -1} \Big( a^{(n)}_{0} \phi ^{0}_{(n)} + a^{(n)}_{0i} \phi ^{0i}_{(n)} +a^{(n)}_{05} \phi ^{05}_{(n)} \Big)
\bigg], 
\end{eqnarray}
where $a^{(0)}_{0}$,  $a^{(0)}_{0i}$, $a^{(n)}_{0}$,  $a^{(n)}_{0i}$ and  $a^{(n)}_{05}$ are Lagrange multipliers enforcing the constraints. For this theory, the fundamental Poisson brackets are given by 
\begin{eqnarray}
&& \{  \Phi _\nu ^{(0)} (x), \Pi ^\mu _{(0)} (z) \} = \delta ^{\mu} _\nu \delta ^{3} (x-z), \quad   \{  B _{\alpha \beta}^{(0)}(x), \Pi ^{\mu \nu} _{(0)} (z) \} = \frac{1}{2}(\delta ^{\mu}    _\alpha \delta ^{\nu} _\beta - \delta ^{\mu}    _\beta  \delta ^{\nu} _\alpha )  \delta ^{3} (x-z), \nonumber \\
&& \{  \Phi _H ^{(l)} (x), \Pi ^L  _{(n)}(z) \} = \delta ^{l} _{n} \delta ^{L} _H \delta ^{3} (x-z), \quad \{  B _{HL} ^{(l)}(x), \Pi ^{M N} _{(n)} (z) \} = \frac{1}{2}\delta ^{l} _{n}(\delta ^{M } _{H}  \delta ^{N } _{L} - \delta ^{M} _{L}  \delta ^{N } _{H}  )  \delta ^{3} (x-z). \quad
\end{eqnarray}
On the other hand, by demanding consistency among the constraints, we find the following secondary constraints 
\begin{eqnarray}
\psi ^{0} _{(n)} &\equiv & \partial _i \Pi ^i _{(n)} + \frac{n}{R} \Pi ^5 _{(n)}  \approx 0 , \nonumber\\
\psi ^{0i}_{(n)} & \equiv & m\Pi ^{i}_{(n)} +2 \partial _j \Pi ^{i j}_{(n)} + \frac{n}{R} 2 \Pi ^{i 5}_{(n)}  \approx  0, \nonumber \\
 \psi ^{05}_{(n)}  & \equiv &  m\Pi ^{5}_{(n)} + 2\partial _j \Pi ^{5 j}_{(n)} \approx  0 .
\end{eqnarray}
For this theory, there are not third constraints.  In this manner, we have found the following set of constraints 
\begin{eqnarray}
\qquad &&\phi ^{0}_{(0)}  \equiv  \Pi ^{0 }_{(0)} \approx 0, \nonumber\\ 
&& \phi ^{0 i  }_{(0)} \equiv \Pi ^{0 i  }_{(0)} \approx 0, \nonumber\\  
&& \psi ^{0} _{(0)} \equiv  \partial _i \Pi ^i _{(0)}\approx 0 ,\nonumber \\
&& \psi ^{0i}_{(0)} \equiv  m\Pi ^{i}_{(0)} +2 \partial _j \Pi ^{i j}_{(0)}  \approx  0,\nonumber\\
&& \phi ^{0}_{(n)}  \equiv  \Pi ^{0}_{(n)} \approx 0,  \nonumber\\
&& \phi ^{0i}_{(n)}  \equiv  \Pi ^{0 i  }_{(n)} \approx 0, \nonumber\\
&& \phi ^{05}_{(n)}  \equiv  \Pi ^{0 5  }_{(n)} \approx 0, \nonumber\\
&& \psi ^{0} _{(n)} \equiv  \partial _i \Pi ^i _{(n)} + \frac{n}{R} \Pi ^5 _{(n)}  \approx 0 , \nonumber\\
&& \psi ^{0i}_{(n)}  \equiv  m\Pi ^{i}_{(n)} +2 \partial _j \Pi ^{i j}_{(n)} + \frac{n}{R} 2 \Pi ^{i 5}_{(n)}  \approx  0, \nonumber \\
&& \psi ^{05}_{(n)}   \equiv   m\Pi ^{5}_{(n)} + 2\partial _j \Pi ^{5 j}_{(n)} \approx  0,
\end{eqnarray}
we are able to observe that all these $10k-2$ constraints are of first class. It is important to comment that the  St{\"{u}}eckelberg's field convert to Proca Kalb-Ramond theory in a  full gauge theory. In the Proca Kalb-Ramond model studied in above section, there are only second class constraints and there are  not reducibility relations among the constraints. In this St{\"{u}}eckelberg Kalb-Ramond theory there are only first class constraints and there are reducibility among the constraints in both, zero modes and $kk$-excitations. These reducibility conditions are given by  the following $k$ relations 
\begin{eqnarray}
&& \partial _{i} \psi ^{ 0 i} _{(0)} - m\psi^{0}  _{(0)} =0,  \nonumber \\ 
&& \partial _{i} \psi ^{ 0 i} _{(n)} - m\psi^{0}  _{(n)} + \frac{n}{R}\psi^{05}  _{(n)}=0,
\end{eqnarray}
thus, the number of independent first class constraints is  $[(10k-2)-k]=9k-2$. In this manner, the counting of physical degrees of freedom is carry out in the following way; there are $30k-10$ dynamical variables and $9k-2$ independent first class constraints, thus, the number of physical degrees of freedom is 
\begin{eqnarray}
DF&=&\frac{1}{2}[ 30k -10 -2(9k-2)]= {\bf 6 k - 3},
\end{eqnarray}
we observe if  $k=1$, then there are   3 degrees of freedom  as expected. In fact,  St{\"{u}}eckelberg Kalb-Ramond and Proca Kalb-Ramond have  the same number of physical degrees of freedom, however,  the former  is a full gauge theory while the latter is not.  We also can observe,   that  for each excitation there is a contribution of $6$ degrees of freedom, just as it  is present in  the Kalb-Ramond theory. \\
We have observed that St{\"{u}}eckelberg Kalb-Ramond  is a reducible system with only  first class constraints, this means that the theory is a gauge theory. Hence, we shall calculate the gauge transformations of the theory. For this aim, we define the following gauge generator 
\begin{eqnarray}
G &=& \int \bigg[ \epsilon ^{(0)}_{0} \phi ^{0}_{(0)} 
+  \epsilon ^{(0)}_{0i} \phi ^{0 i  }_{(0)} 
+  \epsilon ^{(0)}_{} \psi ^{0} _{(0)} 
+\epsilon ^{(0)}_{i} \psi ^{0i}_{(0)} 
+ \epsilon ^{(n)}_{0} \phi ^{0}_{(n)}\nonumber \\
&&  +    \epsilon ^{(n)}_{0i} \phi ^{0i}_{(n)} 
 + \epsilon ^{(n)}_{} \psi ^{0} _{(n)} 
+ \epsilon ^{(n)}_{i} \psi ^{0i}_{(n)} 
+  \epsilon ^{(n)}_{05} \phi ^{05}_{(n)}
+  \epsilon ^{(n)}_{5}  \psi ^{05}_{(n)}\bigg] d^3z.  
\end{eqnarray}
thus, the following gauge transformations of the theory are obtained 
\begin{eqnarray}
&& \Phi _{0 } ^{(0)} \rightarrow  \Phi _{0 } ^{(0)} + \partial _{0} \epsilon  ^{(0)}, \nonumber \\
&& \Phi _{i } ^{(0)} \rightarrow  \Phi _{i } ^{(0)} - \partial _{i} \epsilon  ^{(0)} + m \epsilon _{i}  ^{(0)}, \nonumber \\
&& B _{0i } ^{(0)} \rightarrow  B _{0i } ^{(0)} -  \partial _{0 } \epsilon _{i}  ^{(0)}, \nonumber \\
&&  B _{ij } ^{(0)} \rightarrow  B _{ij } ^{(0)} + \partial _{i } \epsilon _{j}  ^{(0)} - \partial _{j } \epsilon _{i}  ^{(0)} , \nonumber \\
&& \Phi _{0 } ^{(n)} \rightarrow  \Phi _{0 } ^{(n)} + \partial _{0} \epsilon  ^{(n)}, \nonumber \\
&& \Phi _{i } ^{(n)} \rightarrow  \Phi _{i } ^{(n)} - \partial _{i} \epsilon  ^{(n)} +  m \epsilon _{i}  ^{(n)}, \nonumber \\
&& \Phi _{5 } ^{(n)} \rightarrow  \Phi _{5 } ^{(n)} + \frac{n}{R}  \epsilon  ^{(n)} - m  \epsilon _{5} ^{(n)}, \nonumber \\
&& 
B _{0i } ^{(n)} \rightarrow  B _{0i } ^{(n)} -  \partial _{0 } \epsilon _{i}  ^{(n)}, \nonumber \\
&& B _{05 } ^{(n)} \rightarrow  B _{05 } ^{(n)} + \partial _{0 } \epsilon _{5}  ^{(n)}, \nonumber \\
&&  B _{ij } ^{(n)} \rightarrow  B _{ij } ^{(n)} + \partial _{i } \epsilon _{j}  ^{(n)} - \partial _{j } \epsilon _{i}  ^{(n)} , \nonumber \\
&&  B _{i5 } ^{(n)} \rightarrow  B _{i5 } ^{(n)} + \frac{n}{R}  \epsilon _{i} ^{(n)} - \partial _{i}  \epsilon _{5} ^{(n)},
\end{eqnarray}
we can write these gauge transformations in the following compact  form 
\begin{eqnarray}
&& \delta \Phi _{\mu } ^{(0)} = - \partial _{\mu} \epsilon  ^{(0)}  
+ m \epsilon ^{(0)}_{\mu},\nonumber \\
&& \delta B _{\mu \nu  } ^{(0)} =  \partial _{\mu } \epsilon _{\nu}  ^{(0)} - \partial _{\nu  } \epsilon _{\mu}  ^{(0)},\nonumber \\
&& \delta \Phi _{\mu } ^{(n)} = - \partial _{\mu} \epsilon  ^{(n)}  
+ m \epsilon ^{(n)}_{\mu},\nonumber \\
&& \delta \Phi _{5 } ^{(n)} = \frac{n}{R} \epsilon  ^{(n)}  
- m \epsilon ^{(n)}_{5},\nonumber \\
&& \delta B _{\mu \nu  } ^{(n)} =  \partial _{\mu } \epsilon _{\nu}  ^{(n)} - \partial _{\nu  } \epsilon _{\mu}  ^{(n)} , \nonumber \\
&& \delta B _{\mu 5  } ^{(n)} =  \frac{n}{R} \epsilon _{\mu}  ^{(n)} - \partial _{\mu  } \epsilon _{5}  ^{(n)}.
\end{eqnarray}
It is interesting to observe that under the following fixed gauge 
\begin{eqnarray}
&&  \epsilon ^{(n)}_{} = \frac{R}{n}( m \epsilon ^{(n)}_{5}- \Phi _{5} ^{(n)} ),\nonumber \\
&& \epsilon ^{(n)}_{\mu} = \frac{R}{n}( \partial _{\mu} \epsilon ^{(n)}_{5} - B_{\mu 5} ^{(n)} ),
\end{eqnarray}
the fields transform like  
\begin{eqnarray}
&& \delta \Phi _{\mu } ^{(n)} = \frac{R}{n} \partial _{\mu} \Phi _{5}-  B _{\mu 5  } ^{(n)},\nonumber \\
&& \delta B _{\mu \nu  } ^{(n)} =  - \partial _{\mu }  B _{\nu 5  } ^{(n)} +  \partial _{\nu }  B _{\mu 5  } ^{(n)} , 
\end{eqnarray}
under that fixed  gauge the effective Lagrangian (\ref{eq70}) is reduced to 
\begin{eqnarray}
\mathcal{L}_{} &=& \frac{1}{2 \times 3!} H^{(0)}_{\mu \nu \lambda} H^{ \mu \nu \lambda}_{(0)} - \frac{1}{4}(m B ^{(0)}_{\mu \nu  } - \Phi^{(0)} _{\mu \nu })(m B^{\mu \nu  }_{(0)} - \Phi ^{\mu  \nu }_{(0)}) + \sum ^{\infty}_{n=1} \bigg[ \frac{1}{2 \times 3!} H^{(n)}_{\mu \nu \lambda} H^{ \mu \nu \lambda}_{(n)} \nonumber\\
&& - \frac{1}{4}(m B ^{(n)}_{\mu \nu  } - \Phi^{(n)} _{\mu \nu })(m B^{\mu \nu  }_{(n)} - \Phi ^{\mu  \nu }_{(n)})  -\frac{1}{2}\Big( \frac{n}{R}\Big) ^{2}  \Phi^{(n)} _{\mu} \Phi _{(n)} ^{\mu}  +\frac{1}{4}\Big( \frac{n}{R}\Big) ^{2}  B^{(n)} _{\mu \nu }  B^{\mu \nu } _{(n)} \bigg].
\end{eqnarray}
this means that the fields $\Phi^{(n)}_5$ and $B_{5\mu}^{(n)}$ has been absorbed and they are identified as pseudo-Goldstone bosons, something similar is  also present in the free  5D  St{\"{u}}eckelberg  theory \cite{ 17}. \\
On the other hand, because of the zero modes and the $kk$-excitations are not mixed in the constraints we procedure to  calculate the Dirac brackets for all the  modes.  In fact, by fixing the gauge  we have  the following constraints for the zero mode 
\begin{eqnarray}
\qquad && \chi ^{1}_{(0)}  \equiv  \Pi ^{0 }_{(0)}, \quad  \chi ^{2}_{(0)}  \equiv   \Phi ^{(0)} _{0}, \nonumber\\
&&  \chi ^{3}_{(0)}  \equiv  \Pi ^{0 i  }_{(0)}, \quad \chi ^{4}_{(0)}  \equiv  B _{0 i  }^{(0)},  \nonumber\\
&&   \chi ^{5}_{(0)}  \equiv  \partial _i \Pi ^i _{(0)} , \quad  \chi ^{6}_{(0)}  \equiv  \partial ^i \Phi _{i} ^{(0)},  \nonumber\\
&& \chi ^{7}_{(0)}  \equiv m \Pi ^{i}_{(0)} +2 \partial _j \Pi ^{i j}_{(0)}
+ \partial ^{i} p _{(0)} , \quad \chi ^{8}_{(0)}  \equiv \partial ^{j} B _{i j}^{(0)}
+ \partial _{i} q_{(0)},  
\label{eq93}
\end{eqnarray}
 now the constraints are of second class and we have introduced the auxiliary variables $q_{(0)}$ and  $p _{(0)}$  in order to have independent second class constraints. The auxiliary variables satisfy 
\begin{eqnarray}
 \{   q ^{(0)} (x),   p _{(0)} (z ) \}  _{} =  \delta ^{3} (x -z ). 
\end{eqnarray}
In this manner,  the nonzero Poisson brackets among the constraints (\ref{eq93}) are given by 
\begin{eqnarray}
 \{ \chi ^{1}_{(0)} (x), \chi ^{2}_{(0)} (z) \}   & = & %\{ \Pi ^{0 }_{(0)}(x), \Phi ^{(0)} _{0}(z) \}= 
 - \delta ^{3}(x-z),\nonumber \\
 \{ \chi ^{3}_{(0)}(x), \chi ^{4}_{(0)}(z) \}   & = & % \{ \Pi ^{0 i  }_{(0)}(x),     B _{0 j  } ^{(0)} (z)\}=  
 -\frac{1}{2}\delta _{j}^{i} \delta ^{3}(x-z),\nonumber \\
  \{ \chi ^{5}_{(0)}(x), \chi ^{6}_{(0)} (z)\}   & = &  %\{  \partial _i \Pi ^{i} _{(0)}(x), \partial ^{j} \Phi _{j} ^{(0)}(z) \}= 
   - \partial _i \partial ^i \delta ^{3} (x -z),\nonumber \\
 \{ \chi ^{6}_{(0)}(x), \chi ^{7}_{(0)}(z) \}   &=& %\{ \partial ^j \Phi _j ^{(0)} (x),   m \Pi ^{i}_{(0)} (z)\} = 
  m  \partial ^{i} \delta ^{3}(x -z ),\nonumber \\
 \{ \chi ^{7}_{(0)}(x), \chi ^{8}_{(0)}(z) \} &=& % \{ 2 \partial _j \Pi ^{i j}_{(0)}  (x),   \partial ^l B _{k l} ^{(0)} (z) \} + \{  \partial ^{i} p _{(0)} (x),    \partial _{k} q ^{(0)}(z) \}
  -  \delta ^{i}_{k} \partial _{j} \partial ^{j} \delta ^{3}(x -z ),
 \label{eq94}
\end{eqnarray}
thus, we form  the following matrix whose entries are given by the Poisson brackets (\ref{eq94}) 
\begin{eqnarray}
\left(C_{\alpha \beta} ^{(0)}\right)
= \left(\begin{array}{cccccccc}
0 & -1 & 0 & 0 & 0 & 0 & 0 & 0\\
1  & 0 & 0 & 0 & 0 & 0 & 0 & 0\\
0 & 0 & 0 & -\frac{1}{2}\delta ^{i}_{j} & 0 & 0 & 0 & 0\\
0 & 0 & \frac{1}{2}\delta ^{i}_{j} & 0 & 0 & 0 & 0 & 0\\
0 & 0 & 0 & 0 & 0 &  -\nabla ^{2} & 0 & 0\\
0 & 0 & 0 & 0 &    \nabla ^{2}& 0 & m\partial ^{i} & 0\\
0 & 0 & 0 & 0 & 0 &  -m\partial ^{i} & 0 &  -\delta ^{i}_{j} \nabla ^{2}  \\
0 & 0 & 0 & 0 & 0 & 0 & \delta ^{i}_{j} \nabla ^{2} & 0
\end{array} \right)  \delta^{3}(x-z), \nonumber
\end{eqnarray}
the inverse of this matrix is 
\begin{eqnarray}
\left(C^{\alpha \beta} _{(0)}\right)
= \left(\begin{array}{cccccccc}
0 & 1 & 0 & 0 & 0 & 0 & 0 & 0\\
-1  & 0 & 0 & 0 & 0 & 0 & 0 & 0\\
0 & 0 & 0 & 2\delta _{i j} & 0 & 0 & 0 & 0\\
0 & 0 & -2 \delta _{ij} & 0 & 0 & 0 & 0 & 0\\
0 & 0 & 0 & 0 & 0 &  \frac{1}{\nabla ^{2}} & 0 & - \frac{m \partial ^{j}}{(\nabla ^{2})^2}\\
0 & 0 & 0 & 0 &   - \frac{1}{\nabla ^{2}} & 0 & 0 & 0\\
0 & 0 & 0 & 03 & 0 & 0 & 0 &  \frac{ \delta ^{j}_{i}}{\nabla ^{2}}  \\
0 & 0 & 0 & 0 &  \frac{m \partial ^{j}}{(\nabla ^{2})^2} & 0 &  -\frac{ \delta ^{j}_{i}}{\nabla ^{2}} & 0
\end{array} \right)  \delta^{3}(x-z). \nonumber
\end{eqnarray}
In this manner, we obtain the following nonzero Dirac brackets among the physical fields 
\begin{eqnarray}
 \{   \Phi _{0} ^{(0)} (x),   \Pi ^{0}  _{(0)} (z ) \}  _{D}  &=& 2 \delta ^{3} (x -z ), \nonumber\\
\ \{   B _{0i} ^{(0)} (x),   \Pi ^{0j}  _{(0)} (z ) \}  _{D} &=&  \delta ^{j}_{i} \delta ^{3} (x -z ), \nonumber \\ 
 \{   \Phi _{i} ^{(0)} (x),   \Pi ^{j}  _{(0)} (z ) \}  _{D}  &=&  [\delta ^{j}_{i} - \frac{1}{\nabla ^2}  \partial _{i} \partial ^{j}] \delta ^{3} (x  -z ), \nonumber \\
 \{   \Phi _{i} ^{(0)} (x),    \Pi ^{jk}  _{(0)} (z ) \}  _{D}  &=&  \frac{m}{2\nabla ^2} [ \delta ^{j}_{i}   \partial ^{k} - \delta ^{k}_{i} \partial ^{j}  ] \delta ^{3} (x -z), \nonumber \\
 \{   B _{ij} ^{(0)} (x),   \Pi ^{kl}  _{(0)} (z ) \}  _{D}  &=&\frac{1}{2}[ \delta ^{k}_{i}\delta ^{l}_{j} - \delta ^{l}_{i}\delta ^{k}_{j} - \frac{1 }{\nabla ^2} (\delta ^{k}_{i} \partial ^{l} \partial _{j}  - \delta ^{l}_{i} \partial ^{k} \partial _{j} - \delta ^{k}_{j} \partial ^{l} \partial _{i} +  \delta ^{l}_{j} \partial ^{k} \partial _{i})  ]  \delta ^{3} (x  -z ).
\end{eqnarray}
and we can observe that the Dirac brackets among physical  and the auxiliary variables  vanish as expected. In fact, the auxiliary fields do not contribute to the theory and they can be  taken as zero at the end of the calculations \cite{12}. \\
%%%%%%%%%%%%%%%%%%%%%%%%%%%%%%%%%%%%%%%%%%%%%%%%%%%%%%%%%%%%%%%%%%%
Now, we will calculate the Dirac brackets for the $kk$-excitations. In fact, by fixing the gauge and introducing auxiliary variables we obtain the following set of second class constraints 
\begin{eqnarray}
\qquad && \chi ^{1}_{(n)}  \equiv  \Pi _{0 }^{(n)}, \quad  \chi ^{2}_{(n)}  \equiv   \Phi _{(n)} ^{0},  \nonumber\\
&&       \chi ^{3}_{(n)}  \equiv  \Pi ^{0 i  }_{(n)}, \quad \chi ^{4}_{(n)}  \equiv  B _{0 i  }^{(n)},  \nonumber\\
&&    \chi ^{5}_{(n)}  \equiv  \Pi ^{05} _{(n)} , \quad  \chi ^{6}_{(n)}  \equiv B_{05} ^{(n)},\nonumber\\
&&  \chi ^{7}_{(n)}  \equiv  \partial _i \Pi ^{i} _{(n)} + \frac{n}{R}\Pi ^{5} _{(n)}, \quad  \chi ^{8}_{(n)}  \equiv  \partial ^i \Phi _i ^{(n)}, \nonumber\\
&&   \chi ^{9}_{(n)}  \equiv m \Pi ^{i}_{(n)} +2 \partial _j \Pi ^{i j}_{(n)} + \frac{n}{R}2\Pi ^{i5} _{(n)} + \partial ^{i} p _{(n)} , \quad   \chi ^{10}_{(n)}  \equiv \partial ^j B _{i j}^{(n)} +  \partial _{i} q ^{(n)}, \nonumber\\
&&   \chi ^{11}_{(n)}  \equiv m \Pi ^{5}_{(n)} +2 \partial _j \Pi ^{5 j}_{(n)} , \quad \chi ^{12}_{(n)}  \equiv \partial ^j B _{5 j} ^{(n)} ,
 \label{eq98}
\end{eqnarray}
where the auxiliary variables   $q_{(n)}$ y $p _{(n)}$ satisfy    the brackets 
\begin{eqnarray}
 \{   q ^{(n)} (x),   p _{(n)} (z ) \}  _{} =  \delta ^{3} (x -z ). 
\end{eqnarray}
The nonzero Poisson brackets among the constraints (\ref{eq98}) are given by 
\begin{eqnarray}
 \{ \chi ^{1}_{(n)} (x), \chi ^{2}_{(n)} (z) \}   & = & %\{ \Pi ^{0 }_{(n)}(x), \Phi ^{(n)} _{0}(z) \}= 
 - \delta ^{3}(x-z),\nonumber \\
 \{ \chi ^{3}_{(n)}(x), \chi ^{4}_{(n)}(z) \}   & =& % \{ \Pi ^{0 i  }_{(n)}(x),     B _{0 j  } ^{(n)} (z)\}= 
 - \frac{1}{2}\delta _{j}^{i} \delta ^{3}(x-z),\nonumber \\
 \{ \chi ^{5}_{(n)}(x), \chi ^{6}_{(n)}(z) \}   & = & %\{ \Pi ^{0 5 }_{(n)}(x),     B _{0 5  } ^{(n)} (z)\}=  
 - \frac{1}{2} \delta ^{3}(x-z),\nonumber \\
 \{ \chi ^{7}_{(n)}(x), \chi ^{8}_{(n)} (z)\}   & = &  %\{  \partial _i \Pi ^{i} _{(n)}(x), \partial ^{j} \Phi _{j} ^{(n)}(z) \}=  
 - \partial _i \partial ^i \delta ^{3} (x -z),\nonumber \\
 \{ \chi ^{8}_{(n)}(x), \chi ^{9}_{(n)}(z) \}   &=& %\{ \partial ^j \Phi _j ^{(n)} (x),   m \Pi ^{i}_{(n)} (z)\} =  
 m  \partial ^{i} \delta ^{3}(x -z ),\nonumber \\
 \{ \chi ^{9}_{(n)}(x), \chi ^{10}_{(n)}(z) \}  &=& % \{ 2 \partial _j \Pi ^{i j}_{(n)}  (x),   \partial ^l B _{k l} ^{(n)} (z) \} + \{  \partial ^{i} p _{(n)} (x),    \partial _{k} q ^{(n)}(z) \} = 
 -  \delta ^{i}_{k} \partial _{j} \partial ^{j} \delta ^{3}(x -z ), \nonumber \\ 
 \{ \chi ^{9}_{(n)}(x), \chi ^{12}_{(n)}(z) \}  &=& % \{ \frac{n}{R}2\Pi ^{i5} _{(n)} (x),   \partial ^l B _{5 l} ^{(n)} (z) \}
  = \frac{n}{R}\partial ^i \delta ^{3}(x -z ),
  \nonumber \\
 \{ \chi ^{11}_{(n)}(x), \chi ^{12}_{(n)}(z) \}  &=& % \{ 2 \partial _j \Pi ^{5 j}_{(n)}  (x),   \partial ^{l} B _{5 l} ^{(n)} (z) \}  = 
 - \partial _{i} \partial ^{i} \delta ^{3}(x - z),
\end{eqnarray}
by using these brackets we construct the  matrix  
\begin{eqnarray}
\left(C_{\alpha \beta} ^{(n)}\right)
= \left(\begin{array}{cccccccccccc}
0 & -1 & 0 & 0 & 0 & 0 & 0 & 0 & 0 & 0 & 0 & 0\\
1 & 0  & 0 & 0 & 0 & 0 & 0 & 0 & 0 & 0 & 0 & 0\\
0 & 0 & 0 & -\frac{1}{2}\delta ^{i}_{j} & 0 & 0 & 0 & 0 & 0 & 0 & 0 & 0\\
0 & 0 &  \frac{1}{2}\delta ^{i}_{j} & 0 & 0 & 0 & 0 & 0 & 0 & 0 & 0 & 0\\
0 & 0  & 0 & 0 & 0 & -\frac{1}{2} & 0 & 0 & 0 & 0 & 0 & 0\\
0 & 0  & 0 & 0 &  \frac{1}{2} & 0 & 0 & 0 & 0 & 0 & 0 & 0\\
0 & 0  & 0 & 0 & 0 & 0 & 0 & -\nabla ^{2} & 0 & 0 & 0 & 0\\
0 & 0  & 0 & 0 & 0 & 0 &  \nabla ^{2} & 0 &  m\partial ^{i} & 0 & 0 & 0\\
0 & 0  & 0 & 0 & 0 & 0 & 0 &  -m\partial ^{i}  & 0 &  -\delta ^{i}_{j} \nabla ^{2} & 0 & \frac{n}{R}\partial ^i \\
0 & 0  & 0 & 0 & 0 & 0 & 0 & 0 & \delta ^{i}_{j} \nabla ^{2}  & 0 & 0 & 0\\
0 & 0  & 0 & 0 & 0 & 0 & 0 & 0 & 0 & 0 & 0 &  -\nabla ^{2}\\
0 & 0  & 0 & 0 & 0 & 0 & 0 & 0 &  -\frac{n}{R}\partial ^i & 0 &  \nabla ^{2}& 0
\end{array} \right)  \delta^{3}(x-z). \nonumber 
\end{eqnarray}
whose inverse is given by 
\begin{eqnarray}
\left(C^{\alpha \beta} _{(n)}\right)
= \left(\begin{array}{cccccccccccc}
 0 & 1 & 0 & 0 & 0 & 0 & 0 & 0 & 0 & 0 & 0 & 0\\
-1 & 0 & 0 & 0 & 0 & 0 & 0 & 0 & 0 & 0 & 0 & 0\\
0 & 0 & 0 & 2\delta ^{j}_{i} & 0 & 0 & 0 & 0  & 0 & 0 & 0 & 0 \\
0 & 0 & -2 \delta ^{j}_{i} & 0 & 0 & 0 & 0 & 0 & 0 & 0 & 0 & 0 \\
0 & 0 & 0 & 0 & 0 & 2 & 0 & 0 & 0 & 0 & 0 & 0 \\
0 & 0 & 0 & 0 & -2 & 0 & 0 & 0 & 0 & 0 & 0 & 0 \\
0 & 0 & 0 & 0 & 0 & 0 & 0 &  \frac{1}{\nabla ^{2}} & 0 & - \frac{m \partial ^{j}}{(\nabla ^{2})^2}& 0 & 0\\
0 & 0 & 0 & 0 & 0 & 0 & -\frac{1}{\nabla ^{2}} & 0 & 0 & 0 & 0 & 0 \\
0 & 0 & 0 & 0 & 0 & 0 & 0 & 0 & 0 & \frac{ \delta ^{j}_{i}} {\nabla ^{2}} & 0 & 0 \\
0 & 0 & 0 & 0 & 0 & 0 & \frac{m \partial ^{j}}{(\nabla ^{2})^2} & 0 & -\frac{ \delta ^{j}_{i}}{\nabla ^{2}} & 0 & -\frac{n \partial ^{j}}{R(\nabla ^{2})^2} & 0 \\
0 & 0 & 0 & 0 & 0 & 0 & 0 & 0 & 0 & \frac{n \partial ^{j}}{R(\nabla ^{2})^2} & 0 &  \frac{1}{\nabla ^{2}} \\
0 & 0 & 0 & 0 & 0 & 0 & 0 & 0 & 0 & 0 &  -\frac{1}{\nabla ^{2}} & 0 \\
\end{array} \right)  \delta^{3}(x-z). \nonumber
\end{eqnarray}
By using this matrix, we obtain the following nonzero Dirac brackets 
\begin{eqnarray}
 \{   \Phi _{0} ^{(n)} (x),   \Pi ^{0}  _{(n)} (z ) \}  _{D} & =& 2 \delta ^{3} (x -z ),  \nonumber \\
 \{   B _{0i} ^{(n)} (x),   \Pi ^{0j}  _{(n)} (z ) \}  _{D} &= & \delta ^{j}_{i} \delta ^{3} (x -z ), \nonumber \\
 \{   \Phi _{i} ^{(n)} (x),   \Pi ^{j}  _{(n)} (z ) \}  _{D}  &=&  [\delta ^{j}_{i} -\frac{1}{\nabla ^2}  \partial _{i} \partial ^{j}] \delta ^{3} (x  -z ),  \nonumber \\
 \{   \Phi _{i} ^{(n)} (x),    \Pi ^{jk}  _{(n)} (z ) \}  _{D}  &=&  \frac{m}{2\nabla ^2} [\delta ^{j}_{i}   \partial ^{k} - \delta ^{k}_{i} \partial ^{j}  ] \delta ^{3} (x -z), \nonumber \\
 \{   B _{ij} ^{(n)} (x),   \Pi ^{kl}  _{(n)} (z ) \}  _{D}  &=&\frac{1}{2}[ \delta ^{k}_{i}\delta ^{l}_{j} - \delta ^{l}_{i}\delta ^{k}_{j} - \frac{1 }{\nabla ^2} (\delta ^{k}_{i} \partial ^{l} \partial _{j}  - \delta ^{l}_{i} \partial ^{k} \partial _{j} - \delta ^{k}_{j} \partial ^{l} \partial _{i} +  \delta ^{l}_{j} \partial ^{k} \partial _{i})  ]  \delta ^{3} (x  -z ), 
\end{eqnarray}
and the brackets among physical and auxiliary variables vanish. It is important to remark, that all  results of this work   are absent in the literature. \\
\section{Conclussions and Prospects}
In this paper,  the Hamiltonian analysis for a  $5D$ Kalb-Ramond,  $5D$ Proca Kalb-Ramond and  St{\"{u}}eckelberg's  Kalb-Ramond theories with a compact dimension  has been performed. Respect to   $5D$ Kalb-Ramond theory,    we obtained the complete canonical description of the theory. After performing  the  compactification of the fifth dimension on a $S^1/\mathbf{Z_2}$ orbifold, we found that  the  effective theory is composed by  a $4D$ Kalb-Ramond  theory identified with the  zero-mode plus  a tower of  $kk$-excitations. We reported  the complete constraints program,  we found that  the  constraints of the theory are of first class and  reducible.  From the gauge transformations of the theory and by fixing a particular gauge,  we identified a tower of massive  fields and the fields  $B_{5\mu}^{(n)}$ are identified as   pseudo-Goldston bosons. Furthermore,  in order to obtain a irreducible set of constraints we introduced auxiliary variables   and  we calculate   the fundamental  Dirac's   brackets for  the  zero modes   and the  $kk$-excitations.\\
On the other hand,  for the Proca Kalb-Ramond theory we observed  that the theory is not a gauge theory as expected. In fact, for the mode zero and for the $kk$-excitations we found that there are only second class constraints, there are not reducibility conditions and there are not present pseudo-Goldstone bosons. We constructed the Dirac brackets  for the zero mode and the $kk$-excitations. \\
Furthermore, we performed the Hamiltonian analysis for St{\"{u}}eckelberg  Kalb-Ramond theory. We found that the theory have only first class constraints;  there are reducibility conditions among the constraints of the zero mode and reducibility conditions for the $kk$-excitations. By fixing the gauge parameters we can observe that the fields $\Phi^{(n)}_5$ and $B_{5\mu}^{(n)}$ are identified as pseudo-Goldstone bosons, thus, the theory describes a $4D$  St{\"{u}}eckelberg  Kalb-Ramond fields plus a tower of massive $kk$-excitations. In order to construct the Dirac brackets, we used the phase space extension procedure for obtaining a irreducible set of second class constraints and we could  construct  the Dirac brackets for the zero mode and for the $kk$-excitations. In this manner, we have all  tools for performing the quantization of the theories under study. In fact, we can calculate  the  propagators among the physical fields  by using the  Dirac brackets. In this respect,  we would like to  comment that the quantization of the theories   by using the results of this work  and by using the symplectic method is already in progress, and  all these ideas will be the subject of forthcoming works \cite{19}. \\
\noindent \textbf{Acknowledgements}\\[1ex]
This work was supported by CONACyT under Grant No. CB-2010/157641. We would like to thank R. Cartas-Fuentevilla for discussion on the subject and reading the manuscript. \\


\begin{thebibliography}{100}
\bibitem{1}  V.I. Ogievetsky and I.V. Polubarinov, Sov. J. Nucl. Phys. 4 (1967) 156.
\bibitem{2} S. Deser, Phys. Rev. 187 (1969) 1931.
 \bibitem{4} Y. Nambu, Phys. Reports 23 (1976) 250.
\bibitem{4a} D.Z. Freedman and P.K. Townsend, Nucl. Phys. bf B177 (1981) 282.
\bibitem{5a} S. Deser and E. Witten, Nucl. Phys. B178 (1981) 491.
\bibitem{6a} S. Deser, P.K. Townsend and W. Siegel, Nucl. Phys. B184 (1981) 333
\bibitem{5} W. Siegel, Phys. Lett. B, 85, (1979) 333.
\bibitem{6} E.S. Fradkln and M.A. Vasillev, Phys. Lett. B, 85 (1979) 47.
\bibitem{7} E. Cremmer and B. Julia, Nucl. Phys. B, 159 (1979) 141.
\bibitem{8} J. Thierry-Mieg, Y. Ne'eman, Proc. Nat. Acad. Sc. USA 79 (1982) 7068
\bibitem{9} M. Kalb and P. Ramond, Phys. Rev. D9 (1974) 2273.
\bibitem{10} E. Bergshoeff, M. de Roo, B. de Wit and P. van Nieuwenhuizen, Nucl. Phys. B(1982) 97; G. F. Chapline and N. S. Manton, Phys. Lett. B 120 (1983) 105. 
\bibitem{11} B. Julia and G. Toulouse, J. de Phys. 16 (1979) 395.
\bibitem{12}    E. Harikumar, M. Sivakumar, Mod.Phys.Lett. A15 (2000) 121-132. 
\bibitem{13} A. Perez-Lorenzana, J. Phys. Conf. Ser. 18, 224 (2005).
\bibitem{14} A. Muck, A. Pilaftsis and R. Ruckl, Phys. Rev. D 65, 085037 (2002). 
\bibitem{15} H. Novales-Sanchez and J. J. Toscano, Phys. Rev. D, 82, 116012 (2010).
\bibitem{16} E. Stueckelberg, Helv. Phys. Acta 11, 299-312. (1938).
\bibitem{17} Alberto Escalante and Mois\'es Z\'arate, {\it Dirac and Faddeev-Jackiw quantization  of a 5D St{\"{u}}eckelberg theory  with a compact dimension}, submitted to Physical  Review D, (2014). 
\bibitem{19} Alberto Escalante, {\it Faddeev-Jackiw quantization of reducible gauge systems},  in preparation, (2014). 
%\bibitem{3} Boris Kors, Pran Nath, JHEP 0507:069, (2005)
%\bibitem{4} W. Pauli, Rev. Mod. Phys. 13, 203-232. (1941).
%\bibitem{5} H. Ruegg, M. Ruiz-Altaba, Int. J. Mod.Phys. A 19:3265-3348, (2004)
%\bibitem{6} A. H. Chamseddine, Phys. Rev. D 24  3065 (1981); E. Witten,  Phys. Lett. B 155, 15, (1985); C. P. Burgess, A. Font and F. Quevedo, Nucl. Phys. B 272,  661, (1986); S. Ferrara, C. Kounnas and M. Porrati,  Phys. Lett. B 181, 263, (1986) .
%\bibitem{7} C. Marshall, and P. Ramond,  Nucl. Phys. B85, 375-414, (1975).
%\bibitem{8} M. Kalb  and P. Ramond, Phys. Rev. D9, 2273-2284, (1974).
%\bibitem{9} T. J. Allen, M. J. Bowick and A. Lahiri,  Mod. Phys. Lett. A 6,  559, (1991).
%\bibitem{10} M. Gogberashvili,  A. Herrera-Aguilar, D. Malag\'on-Morej\'on and R. R. Mora-Luna, Phys.Lett. B 725,  208-211, (2013). 
%\bibitem{11} G. Weiglein et al. (Physics Interplay of the LHC and the ILC), arXiv:hep-ph/0410364.
%\bibitem{12} M. Henneaux and C. Teitelboim, Quantisation of gauge systems, Princeton University Press, Princeton,(1992).
%\bibitem{13} A. Escalante and L. Carbajal, Ann. Phys. 326, 323179, (2011).
%\bibitem{14} A. Escalante and I. Ruvalcaba-Garc{\'i}a, Int. J. Geom. Methods Mod. Phys. 9, 1250053(2012).
%\bibitem{15}  A. Escalante and J. Manuel-Cabrera, Ann.  Phys. 343,  271¤739, (2014).
%\bibitem{16} L. Faddeev and R. Jackiw, Phys. Rev. Lett. 60, 1692 (1988).

%\bibitem{20}  L. Liao and  Y.  C.  Huang, Ann. Phys.  322, 24691¤72484, (2007).


%\bibitem{sund} Sundermeyer K. { \it Constrained Dynamics}, Lecture Notes in Physics vol.169, Spinger-Verlag, Berlin Heidelberg New York, 1982.
%\bibitem{stueck}E.~C.~G.~Stueckelberg, Helv. Phys. Acta. {\bf 11}, 225 (1938);
%V.~I.~Ogievetskii and I.~V.~Polubarinov, JETP {\bf 14}, 179 (1962).
%\bibitem{higgs} P.~W.~Higgs, Phys. Lett. {\bf 12}, 132 (1964);
%\bibitem{ext}B. K{\"o}rs and P. Nath, arXiv:0503208v2.
%\bibitem{bar}R. Amorim and J. Barcelos-Neto, arxiv:0108171v2.

%\bibitem{3} A. P\'erez-Lorenzana, J. Phys. Conf. Ser. 18, 224 (2005).
%\bibitem{10} A. Escalante, J. Berra, Int.J.Pure Appl. Math. 79,  405-423, (2012). 
%\bibitem{11} J. Govaerts, { \it Topological quantum field theory and pure Yang-Mills dynamics}, in
%Proc. Third Int. Workshop on Contemporary Problems in Mathematical Physics (COPROMAPH3 ), Cotonou (Republic of Benin), 1ï¿½November 2003; Bruno Bertrand and Jan Govaerts,  J.Phys. A 40  F979-F986 (2007); Bruno Bertrand  and  Jan Govaerts, J.Phys. A 40,  9609-9634,  (2007). 
%\bibitem{yo} Alberto Escalante and Moises Zarate {\it A pure Dirac's analysis for a four dimensional  BF-like theory  with a compact dimension}, submitted to international Journal of geometric Methods in physics, (2013).  
%\bibitem{5a} N. Banerjee, R. Banerjee, Mod.Phys.Lett. A11, 1919-1928, (1996).
%\bibitem{12} E. S. Fradkin and G. A. Vilkovisky, Phys. Lett. B55,  224 M, (1975).  
%\bibitem{13} Henneaux, Phys. Rep. C126 (1985) 1.
%\bibitem{1} Th. Kaluza, Sitzungober. Preuss. Akad. Wiss. Berlin (1921) 966;
%O. Klein, Z. Phys. 37 (1926) 895.
%\bibitem{2} M. B. Green, J. H. Schwarz and E. Witten, Superstring Theory (Cambridge University
%Press, Cambridge, 1986); J. Polchinski, String Theory (Cambridge University Press, Cambridge, 1998);  S. T. Yau (ed.), Mathemathical Aspects of String Theory (World Scientific, Singapore, 17 1987).
%\bibitem{5b} Soon-Tae Hong, Yong-Wan Kim, Young-Jai Park and  K.D. Rothe, Mod.Phys.Lett. A17,  435-452, (2002).
%\bibitem{7} H. Novales-Sanchez and J. J. Toscano, Phys. Rev. D, 82, 116012 (2010)
%\bibitem{1} S. Weinberg, The Quantum Theory of Fields (Cambridge University Press, Cambridge, England, 1996), Vols. I-II.
%\bibitem{2}   D. M. Gitman and I.V.Tyutin,  Quantization of fields with constraints, ( Berlin, Germany: Springer. (Springer series in nuclear and particle physics, (1990));  
%A. Hanson, T. Regge and C. Teitelboim. Constrained Hamiltonian Systems (Accademia Nazionale dei Lincei, Roma, (1978)).
%\bibitem{3} Merced Montesinos, G. F. Torres del Castillo, Phys. Rev. A 70:032104, (2004).
%\bibitem{4} M. Mondragon and M. Montesinos, J. Math. Phys. 47, 022301 (2006).
%\bibitem{5} Alberto Escalante. Phys. Lett.B 676:105-111, (2009).
%\bibitem{6} A. Escalante and J. Angel L\'opez-Osio, {\it Hamiltonian analysis for topological and Yang-Mills theories
%expressed as a constrained BF-like theory}, IJPAM  75, 339-352, (2012).
%\bibitem{7} Torres del Castillo G F and Acosta Avalos D, Rev. Mex. Fè¦ç«s. 40 405, (1994).
%\bibitem{8} R. Rosas-Rodriguez, J.Phys.Conf.Ser.24:231-235, (2005).
%\bibitem{9}  J. Barcelos-Neto, T.G. Dargam,  Z.Phys.C67:701-706, (1995). 
%\bibitem{10} R. Rosas-Rodriguez, Int. J. Mod. Phys. A,  23: 895-908, (2008).
%\bibitem{6} A. Muck, A. Pilaftsis and R. Ruckl, Phys. Rev. D 65, 085037 (2002).
%\bibitem{6a} I. Antoniadis, Phys. Lett. B 246, 377, (1990). 
%\bibitem{6b}  J.D. Lykken, Phys. Rev. D 54, 3693 (1996).
%\bibitem{6d} K.R. Dienes, E. Dudas, and T. Gherghetta, Phys. Lett. B 436, 55 (1998); Nucl. Phys. B537, 47 (1999).
%\bibitem{5}  L. Nilse, hep-ph/0601015. 
\end{thebibliography}
\end{document}